# Channel saturation and conductance quantization in single-atom gold constrictions

Jason N. Armstrong, R. M. Schaub, Susan Z. Hua, and Harsh Deep Chopra*

*Laboratory for Quantum Devices, Materials Program, Mechanical and Aerospace Engineering Department, State University of New York at Buffalo, Buffalo, NY, USA 14260*

## Abstract

Notwithstanding the discreteness of metallic constrictions, it is shown that the finite elasticity of stable, single-atom gold constrictions allows for a continuous and reversible change in conductance, thereby enabling observation of channel saturation and conductance quantization. The observed channel saturation and signature for conductance quantization is achieved by superposition of atomic/subatomic-scale oscillations on a retracting/approaching gold tip against a gold substrate of a scanning probe. Results also show that conductance histograms are neither suitable for evaluating the stability of atomic configurations through peak positions or peak height nor appropriate for assessing conductance quantization. A large number of atomic configurations with similar conductance values give rise to peaks in the conductance histogram. The positions of the peaks and counts at each peak can be varied by changing the conditions under which the histograms are made. Histogram counts below $1G_o$ cannot necessarily be assumed to arise from single-atom constrictions.

## I. INTRODUCTION

Realization of electronic devices based on atomic-sized metal or metal-molecule hybrid systems[1-4] requires a fundamental understanding of conditions under which conductance quantization prevails,[5] their electronic structure and regulation of electron transmission behavior,[6-19] and their mechanical stability against disruptive forces of entropic thermal fluctuations.[20-23] In addition, positioning and manipulation of individual atoms requires fundamental understanding of the operative forces,[24-27] and the ability to measure them.[28-37] Whereas quantization of conductance was first shown experimentally using two-dimensional electron gas,[38-39] its demonstration in atomic-sized *metal* constrictions has been stymied by the difficulty of *continuously* varying the diameter of the constriction on the scale of the Fermi wavelength (which is on the same order of inter-atomic distances in metals) without causing a concomitant atomic rearrangement within the constriction. The signature of conductance quantization in metal point contacts would be relatively straightforward if stepwise changes in conductance only occurred at integer multiples of $2e^2/h$ ($= 1G_o \approx 77.48\ \mu S$), the quantum of conductance, as the diameter of the constriction is mechanically increased or decreased; $e$ is the quantum of charge and $h$ is the Planck constant. Instead, the interdependence between atomic (re)arrangement and observed conductance not only give rise to stepwise change in conductance at integer multiples of $G_o$, but also at non- integer values of $G_o$. To illustrate, Fig. 1(a-c) shows example conductance traces for gold, as the constrictions are being broken down to the last atom (experimental details are given later). In these example traces, the last few conductance plateaus mainly take integer or near-integer values of $G_o$ prior to being ruptured. This could be construed as proof of conductance quantization.[40] However Fig. 2(a-c) shows other observed traces for gold where plateaus can be seen at various non-integer values of $G_o$. In other words, although the change in conductance is discrete (or stepwise), it is not quantized. The observation of non-integer conductance plateaus is not accidental either. They occur frequently indicating that the system does not show any particular preference to assume conductance values only at integer multiples of $G_o$.

In an attempt to extract the signature for conductance quantization from such experiments, what is referred to as the conductance histogram was introduced,[41-42] which plots the cumulative counts at a given value of conductance over repeated experiments; conductance histograms were initially used to study the effect of electronic structure on observed conductance in various elements.[43] Figure 3 shows a series of gold conductance histograms, where successive histograms represent the collective total of all preceding experimental cycles. The inset in Fig. 3 shows a zoom-in view of the first three peaks. Distinct peaks can be seen to occur at or close to integer multiples of $G_o$ (dotted vertical lines in the inset of Fig. 3). Moreover, except for an increase in the number of counts with additional experimental cycles, the position of peaks remains unchanged. From such histograms, it has been posited that even though the observed values of conductance may assume integer and non-integer values of $G_o$, the fact that peaks only occur at or near integers of $G_o$ constitutes or at least strongly suggests proof of conductance quantization.[41-42] This is the principal argument for using histograms as a validation tool for conductance quantization in metals. In addition, a characteristic sequence of peaks for sodium conductance histograms (peaks at 1, 3, 5, and $6G_o$, with peaks at 2, 4 and 7 being absent) has been reported and cited as a signature of conductance quantization.[44] However, a small but well-defined peak is observed at $2G_o$ in both sodium and potassium histograms.[45] In addition, two of the authors in Ref. [44] subsequently questioned the validity of using conductance histograms as a proof for conductance quantization, instead interpreting the peaks as arising from preferred atomic reconfigurations.[46] Nonetheless, these authors noted that their original interpretation of the characteristic sequence in sodium conductance histogram in Ref. [44] as a signature of conductance quantization would be "difficult to explain by other models", while adding that the role of preferred atomic rearrangements has to be considered in interpreting conductance histograms.[46]

Since the number of valence electrons has been shown to equal the number of available channels,[47] a one-atom constriction of monovalent gold has a single available channel for conductance. Therefore, a histogram of a true single-atom gold constriction should have a precipitous drop in counts at values exceeding $1G_o$. Yet, an examination of conductance histograms in Fig. 3 shows a continuous and significant number of counts at values just over $1G_o$. This is indicative of contribution to counts in the first peak from contacts larger than a single atom; although, to-date, there have been no means of separating these contributions other than by direct observation in a transmission electron microscope (TEM).[48-49] The possibility of different sized contacts having similar conductance values also cannot be excluded. The counts in the histogram change continuously from one peak to another, indicating that over repeated experimental cycles the system assumes all intermediate states. In this regard, previous studies have shown that these intermediate counts arise from fluctuations between metastable atomic configurations having slightly different transmission probabilities and transitions between different configurations manifesting as random telegraphic noise (RTN).[50]

Naturally, a focal point of discussion on what constitutes proof of conductance quantization has centered on interpreting and understanding the origins of the observed peaks in conductance histograms. As noted earlier, an alternate viewpoint is that peaks in conductance histograms represent favorable or stable atomic configurations, arising from electronic and shell effects.[44, 51-

[53] Force measurements on gold constrictions have also been performed simultaneously with conductance measurements, which show the correlation between mechanical rearrangements of the constriction and the observed conductance.[36-37] These results found that a stepwise change in conductance was always accompanied by a stepwise change in force, indicating the lockstep process of atomic rearrangement of the constriction with a stepwise change in conductance.[54-55] In addition, elongation of gold contacts in multiples of distance that are consistent with slip on {111} closed-packed planes of gold have been observed.[35] Simultaneous measurement of conductance and force during the rupture of a gold constriction is illustrated in Fig. 4(a-b), which shows that the stepwise change in conductance is mirrored by a stepwise change in the measured force as the constriction is being pulled apart from an initially very large size ($\sim 27G_o$, where quantized effects are not expected) down to a single atom whose length scale is pertinent to observation of quantized conductance. More recently, peaks in conductance histograms have been theoretically revisited using molecular dynamic simulations and conductance calculations based on tight binding model.[56] These results show that the minimum cross sectional area of the contacts is not the only factor, but that contact geometry and disorder also plays an important role. The authors noted that a consequence of this interplay is that contacts of different radii may have the same conductance. For the conditions used, their conclusion was that the peak at $1G_o$ is a result of a single-atom or atomic chain.

The data shown in Fig. 1-4 is not just meant as an aid to literature survey, and its subtleties are discussed in further detail in the following. There are pitfalls to assigning the measured conductance value (for example, $1G_o$) to a contact of a given size (for example, a one-atom constriction), and has to be ascertained. The interpretation of what constitutes conductance quantization is far more nuanced. The finite elasticity and stability of particular atomic configurations provides an avenue for extracting the signature for conductance quantization, and these issues are discussed in detail in the following.

## II. EXPERIMENTAL DETAILS

Gold films (200 nm thick) were deposited on a silicon substrate using magnetron sputtering at Ar partial pressure of 3 mtorr in an ultra-high vacuum chamber whose base pressure is ~$10^{-8}$-$10^{-9}$ torr. The gold target used for sputtering was 99.999% pure. A modified atomic force microscope (Ambios Q-Scope Nomad) was used for force-elongation measurements at room temperature. The modified AFM assembly consisted of a dual piezo configuration to control the x-, y-, and z-position of the substrate relative to the cantilever tip. The use of dual piezos for coarse and fine alignment of the substrate relative to the cantilever tip provides low noise and high resolution. The minimum step size of the dual piezo assembly was 4 picometer and the noise was ~5 picometer. The cantilevers used for force measurements had a spring constant of ~11 N/m (Veeco Probes). Apart from the rated spring constant provided by the vendor, the cantilever tips were independently calibrated.[57] The photo-detector was calibrated using the well established optical deflection technique (ODT). The cantilever tip was coated with 60 nm thick gold film. During deposition the cantilever was periodically rotated relative to the sputtering gun to enhance uniformity of the coating on the cantilever tip. While simultaneous measurements of conductance and the force-elongation curves were made using the above described apparatus, all other measurements were made using a custom built, inert atmosphere STM assembly at room temperature. The STM assembly also utilizes the dual piezo configuration. The gold STM tips were prepared by electrochemical etching of an annealed 250 μm diameter gold wire of 99.999% purity. The sharpness of the STM tip is critical for controlled formation of stable atomic sized point contacts, and the process used to prepare the tips was carefully optimized. The electrolyte used is a mixture of 50% hydrochloric acid and 50% ethanol at an etching voltage of 2.4 V. Both AFM and STM assemblies were mounted on a three-stage vibration isolation system for enhanced stability of the constrictions and to minimize the effect of any spurious mechanical vibrations. The probe chamber for STM and AFM measurements as well as the vibration isolation system was enclosed in an acoustic chamber and a Faraday cage. Separately, the entire electronics and data acquisition hardware were also enclosed in another Faraday cage to minimize electrical interference. All AFM and STM measurements as well as tip mounting were done at room temperature in an inert atmosphere (Ar). The contacts were formed by indenting the STM tips against a 200 nm thick gold film sputter-deposited on a silicon substrate. The STM tip is capable of translation along the x-, y-, and z-axes relative to the substrate, with a minimum step size of 4 pm and measured noise of ~5 pm. Apart from the vendor specifications (Physik Instrumente, GmbH), the fine piezo was calibrated using an absolute method, *viz*., forming a parallel-plate capacitor between the tip and substrate holder plates. Amplitudes of superimposed oscillations on the tip-substrate assembly ranged from 0.05 nm to 0.5 nm (peak-to-peak) at frequencies ranging from 0.1 to 100 Hz, with most experiments done at 3 Hz.

The data acquisition hardware consists of a real-time controller, a field-programmable gate array (FPGA), and a data acquisition system. The FPGA is a reprogrammable silicon chip embedded in the controller, that once programmed, is capable of running a large number of parallel operations at very high speeds. The host computer, controller, and FPGA interact to control the probe assembly, each of which requires their own (LabView) program. The host computer passes variables to the real-time controller which can then pass values to the FPGA. In this architecture, the controller is only being used as a 'middle man' between the host computer and the FPGA. There are two output modules (National Instruments 9263) each having 4 analog output channels; four channels from one module are used to send voltage signals to the four piezo voltage amplifiers and one channel is used from the second module to control the bias voltage. There is also an input module (National Instruments 9215) that reads the actual bias voltage coming from the bias voltage signal and the voltage signal coming from the current amplifier. These two signals are measured by the FPGA to compute the conductance of the point contact, which is used as a set point during various modes of operation. The input module (National Instruments 9215) has four separate A/D convertors that simultaneously acquire the conductance, bias voltage, and force (for AFM).

For conductance histograms, the traces were recorded while the contact was being broken at a specified retraction rate and a bias voltage of 250 mV (this bias voltage is used throughout the rest of this work unless otherwise specified). The bin size of the conductance histograms is $0.02G_o$. The program is capable of automatically building the histograms as the traces are being acquired. While the conductance signal is being acquired, the standard deviation can be automatically calculated and written to a separate file at a rate of 20 Hz; 1000 conductance data points (50 milliseconds at 20 kHz) are used for each standard deviation data point. The discrete steps in the conductance signal are due to atomic (structural) rearrangements that lead to a change in conductance. As discussed later, the sections of data that correspond to these transition regions were manually eliminated since they give a spurious contribution to the standard deviation. Only plateaus that exist for at least 200 ms without discrete steps are counted to ensure that the configuration is stable. For noise analysis, the point contacts were formed at a speed of 1 nm/s until the size was greater than $20G_o$ so each point contact is guaranteed to have a new geometry. The contact was then retracted at a rate of 0.25 nm/s until the conductance was $10G_o$ and then retracted at a rate of 3.5 pm/sec until the contact was broken for the actual noise measurements. The retraction rate of 3.5 pm/s was the slowest rate that could consistently break the contact from a size of approximately 10 atoms. The slowest possible retraction rate was chosen to minimize the effect of strain; at even slower retraction rates the diffusion of atoms into the constriction caused the contact to grow faster than it could consistently be broken.

## III. RESULTS AND DISCUSSION

Consider in detail the characteristics of the conductance trace shown in Fig. 2(c). The long plateau at $\sim 1G_o$ stretches by 0.305 nm prior to rupturing. A zoom-in view of this conductance trace along with its simultaneously measured force-elongation curve is shown in Fig. 5(a) and 5(b), respectively. Several observations can be made from Fig. 5. Notice that the short-duration telegraphic conductance fluctuation around $1.54G_o$ in Fig. 5(a) is mirrored by a simultaneous change in force in Fig. 5(b), revealing that the origins of such random telegraphic noise (RTN) is atomic reconfiguration of the constriction; force changes associated with RTN are discussed in detail later. Next, note that the long plateau at $\sim 1G_o$ in fact consists of (at least) two distinct atomic configurations, as seen from the stepwise change in force at the point of (small) stepwise change in conductance from $1.17G_o$ to $1.02G_o$. This highlights the fact that a small stepwise change in conductance does not necessarily mean a small jump in force, the latter being a function of the spring constant for the rupturing atomic configuration, its elasticity, and the state into which it transitions. Moreover, the force-deflection curve in Fig. 5(b) only tells that an atomic reconfiguration has occurred, not the number of atoms involved in the reconfiguration. In fact, the inset in Fig. 5(a) shows that the conductance plateau remains above $1G_o$, except for a brief dip. Yet, counts from such a plateau would be routinely added to the profile of the first peak in a conductance histogram without distinction.

Next, consider the conductance trace shown earlier in Fig. 1(c). A zoom-in view of this trace is shown in Fig. 6(a) along with its simultaneously measured force-deflection curve in Fig. 6(b). Figure 6(a) shows that the long plateau at $\sim 1G_o$ in fact undergoes several atomic reconfigurations. The first atomic reconfiguration is signaled by the large stepwise change in force corresponding to a small but finite stepwise change in conductance from $1.05G_o$ to $0.99G_o$. Again, notice that a small change in conductance does not necessarily mean a small jump in force. A series of atomic reconfigurations occurs towards the end of the plateau, which results in a telegraphic fluctuation followed by the rupture of the contact. At first sight, this elongated plateau at $\sim 1G_o$, which extends by ~0.62 nm prior to rupturing {see note in Ref. [58]}, might be construed as the process of chain formation. However, it cannot even be ascertained whether the plateaus at $1.05G_o$, $0.99G_o$, and $0.84G_o$ corresponds to a single-atom constriction, let alone any suggestion of chain formation. As noted earlier, the force-deflection curves only tell that an atomic reconfiguration has occurred, and not the number of atoms involved in these reconfigurations. Although $1G_o$ is the maximum threshold for a gold single-atom constriction, the reverse is not true. A two-atom constriction is later shown to give rise to net conductance close to or even below $1G_o$, and in fact this has been found to occur frequently. Though in terms of a conductance histogram, the counts from such a long plateau at $\sim 1G_o$ are ordinarily added to the first peak at $\sim 1G_o$.

Next, consider a conductance trace whose last conductance plateau is consistently below $1G_o$. Figure 7(a) shows such a trace along with its simultaneously measured force-elongation curve in Fig. 7(b). The inset in Fig. 7(a) shows a zoom-in view of the plateau at $\sim 1G_o$ and its corresponding force deflection curve. Figure 7 shows that a telegraphic change in conductance occurs within the last plateau that triggers a corresponding change in the measured force, while this atomic rearrangement causes only a minimal change in conductance (from $\sim 0.97G_o$ to $\sim 0.95G_o$). The telegraphic change in conductance occurs when the constriction has been stretched by 0.31 nm, following which the constriction elongates by 0.115 nm before rupturing, for a total elongation of 0.425 nm.[59] Figure 8 shows another example where the last conductance plateau is consistently below $1G_o$, along with its simultaneously measured force-elongation curve in Fig. 8(b). The conductance trace in Fig. 8(a) can be seen to undergo multiple atomic reconfigurations prior to rupturing, as reflected in the corresponding jumps in the measured force in Fig. 8(b). In the examples shown in Fig. 7 and 8, it is again stressed that it cannot be assumed that the last conductance plateau corresponds to a single-atom constriction. While such traces *could be* likely candidates for further scrutinizing the formation of atomic chains, *a priori* it cannot be assumed; TEM offers one route for ascertaining the constriction diameter.[48-49,60] Moreover, without such certainty, no such discretion can be exercised in building a conductance histogram corresponding to a single atom constriction.

Figure 9 and 10 shows examples of conductance traces that undergo random telegraphic noise in conductance during the process of rupturing the constrictions, along with their simultaneously measured force-elongation curves. The window in Fig. 9 shows that the conductance telegraphically fluctuates back and forth between approximately $5G_o$ and $6G_o$ upon stretching. Figure 9(b) shows that the origin of these fluctuations is atomic rearrangement of the constriction between two metastable configurations. The force-elongation curve in Fig. 9(b) shows a finite restoring force for the two configurations. In contrast, for the example shown in Fig. 10, notice the short duration RTN just above $1G_o$, which causes a local change in the slope of the measured force (force constant), giving an apparently reduced force constant.

Figure 11(a) and Fig. 11(b) shows two examples of RTN with the piezo being stationary (no retraction or extension of the piezo). In these examples, thermal fluctuations cause instability in the constrictions, which manifest as RTN, and the cantilever detects the resulting change in force as the system transition between different levels. Figure 11(a) represents the simpler case of two-level fluctuations, whereas Figure 11(b) represents more complicated multi-level fluctuations; see Ref. [50] for detailed discussion of RTN in metal constrictions. The dotted vertical lines in Fig. 11(a), and the highlighted windows in Fig. 11(b) shows the correlation between change in conductance and measured force. In other words, the origin of RTN induced by retraction of the piezo (for example, Figs. 9 and 10) as well as those induced by thermal fluctuations (Fig. 11) are a result of atomic reconfiguration of the constrictions. For other origins of RTN, see Ref. [6, 16, 61-69]. In particular, earlier work on larger diameter metal nanocontacts (where quantized conductance is not dominant) by Ralls and Buhrman[68-69] was attributed to the fluctuation of metastable defects between discrete configurations. In contrast, in the present study, as well as in a recent study[50], mechanical or thermal perturbations causes the metastable contact geometry to fluctuate between discrete configurations leading to slightly different transmission probabilities, and manifesting as RTN. Non-mechanical origins of RTN is beyond the scope of the present study, and will be discussed in the future.

From above discussion, the following observations are made:

**Observation-1:** A small stepwise jump in conductance does not necessarily mean a small stepwise change in force. The magnitude of the force jump is a function of the force constant of the atomic configuration being ruptured, its elasticity, and the state into which it transitions.

**Observation-2:** Conductance traces below $1G_o$ cannot necessarily be assumed to arise from a single-atom constriction from an inspection of conductance traces alone. Although $1G_o$ is the maximum threshold for a gold single-atom constriction, the reverse is not true. A two-atom constriction could also give rise to a net conductance close to or even below $1G_o$.

**Observation-3:** Force-deflection curves only tell that an atomic reconfiguration has occurred, not the number of atoms involved in the reconfiguration.

**Observation-4:** Adding counts to the first peak in conductance histogram from traces at $\sim 1G_o$ does not equate to a single-atom constriction or proof of conductance quantization. Additionally, it does not necessarily indicate a particular atomic configuration of high stability. A peak could equally likely be the result of superposition of counts from different atomic configurations. The monotony of scrutinizing each trace for its contact size cannot be substituted by non-discriminate statistics on a dataset, however large.

**Observation-5:** A simple inspection of plateaus at $\sim 1G_o$ is not a proof of chain formation, unless ascertained by other means (such as TEM).

**Observation-6:** There exists a multitude of atomic configurations that give rise to quasi-continuous values of conductance in a histogram.

It is clear from the conductance traces and load-deflection curves that there exists a multitude of atomic configurations. Compounding this problem is the fact that different atomic configurations may have the same conductance value. The number of counts at a given conductance value in a histogram makes no such distinction. Yet, peaks in histograms have been attributed to either conductance quantization,[41-42, 44] or more recently, the existence of particularly stable atomic configurations.[44, 46, 51-53] For the latter, an assessment of relative stabilities of atomic configurations cannot be obtained from a histogram and requires a mechanistic understanding of how various atomic configurations responds to a perturbation. Description of mechanical stability of an atomic configuration would require knowledge of both the spring constant $k$ and elongation $x$; within the energy landscape, the change in elastic potential energy of an atomic configuration being $\Delta U = \int_{x^i}^{x_f} kx dx$. Therefore, a question that naturally arises is: how are various atomic configurations represented or assimilated in a conductance histogram? One avenue is to analyze the characteristic noise of each conductance plateau, corresponding to each atomic configuration. However, prior to making such analysis, artifacts arising from stepwise segments within a given plateau (for example, those arising from RTN) must be eliminated. This is not only because RTN represents transitions between different atomic configurations (as shown in Fig. 9-11), but also because if such stepwise segments are included, the resulting analysis will give spuriously high noise. This is illustrated in Fig. 12(a) with an example, where the constriction is seen to undergo RTN between two distinct atomic configurations (labeled as Level-1 and Level-2). The inset in Fig. 12(a) shows a zoom-in view of the two levels. The highlighted schematic (green) bins in the inset show the sections of the trace containing the stepwise jumps between the two levels. Figure 12(b) is the conductance histogram of the entire trace in Fig. 12(a), which shows that the conductance either lies in Level-1 or in Level-2, *with no counts in between*. Figure 12(c) is the standard deviation of *all the bins* from the trace in Fig. 12(a). The data points within the red ellipse (on the left) and blue ellipse (on the right) in Fig. 12(c) correspond to the true standard deviation of Level-1 and Level-2, respectively. All other data points in Fig. 12(c) arise from bins that contain the stepwise change between Level-1 and Level-2. These data points not only give rise to spurious data points at conductance values that do not exist in the histogram shown in Fig. 12(b), but also give rise to spuriously high standard deviation for either level (for example, points enclosed in the dotted rectangle). This is the reason why all stepwise changes in conductance have to be eliminated in noise analysis, however laborious. (This type of artifact in noise can be easily demonstrated by binning a square wave from a function generator). In this manner, every stepwise change in all recorded conductance traces was manually eliminated, and the resulting plot is shown in Fig. 13(a). The striking feature of the 'dust map' in Fig. 13(a) is the quasi-continuous distribution of conductance values assumed by various plateaus. Figure 13(b) is a zoom-in view of the dust map in the vicinity of $1G_o$. The red curve in Fig. 13(b) is the average standard deviation of the dust map, whereas the lower green curve roughly outlines the lower bound of the standard deviation. Notice the existence of numerous local minima both above and below $1G_o$. The force-elongation curves shown earlier corroborate the existence of these states above and below $1G_o$. Also note that even with thousands of data points in the dust map, there is a distinct lower bound for the standard deviation. Figure 14(a-b) juxtaposes, respectively, the average value of the standard deviation [obtained from the dust map in Fig. 13(a)] with a histogram shown earlier in Fig. 3. The general minima of the average value of the standard deviation matches well with the peaks in a conventional conductance histogram. Additionally, Fig. 14(a) reveals numerous local minima at far higher resolution. Although the dust map provides information on the quasi-continuous nature of atomic configurations at a higher resolution than a conventional histogram, it still does not provide any information on the number of atoms in each configuration. For example, the local minima just above $1G_o$ could not arise from a single-atom constriction. At the same time, the local minima at and below $1G_o$ cannot be unambiguously attributed only to single atom constrictions.

The conductance of a single-atom gold constriction would saturate at $1G_o$ under a perturbation and would constitute an unambiguous signature of conductance quantization. In this regard, let us consider the effect of a mechanical perturbation in the form of a small mechanical oscillation of subatomic or atomic-scale amplitude that is superimposed on the retracting gold tip, as shown schematically in the inset of Fig. 15(a). The resulting histograms at different (peak-to-peak) amplitudes are shown in Fig. 15(a). For reference, the topmost trace in Fig. 15(a) is the histogram without any oscillation (from Fig. 3). Figure 15(a) shows that with increasing amplitude of oscillation the peak height decreases and the peaks shift to higher conductance values. In other words, peak position is not an intrinsic feature of the metal system but depends on experimental conditions. Figure 15(b) shows a zoom-in view of peak shift around $1G_o$ and is used in the following to discuss the efficacy of applying a small oscillating strain in distinguishing counts arising from a single atom versus multiple atoms and conductance quantization effects.

Figure 16(a) shows a prototypical example (representing the behavior of thousands of similarly behaving contacts) of a two-atom contact that has net conductance close to $1G_o$. Figure 16(a) shows that the conductance changes continuously and reversibly below and above $1G_o$ as the tip is elongated and retracted, and such traces cannot be attributed to a single-atom contact. Equally significant is the fact that depending on the strain state of the system, such a configuration could take a range of conductance values both above and below $1G_o$ over repeated experiments. In a conventional histogram, counts from such a contact would be assigned to the first peak at $1G_o$ without any recourse to their distinction from true single-atom contacts, which highlights the futility of using the conventional conductance histograms.

In contrast, a prototypical example of a conductance trace for a true single-atom contact is shown in Fig. 16(b) as the tip is elongated and retracted. Figure 16(b) shows the hallmark of conductance quantization, *viz*., saturation at $1G_o$ due to the availability of only a single channel for conductance across a single gold atom. Figure 16(c) shows the reversible nature of change in conductance with repeated piezo cycling of the contact corresponding to Fig. 16(a), whereas Fig. 16(d) shows the same for the contact corresponding to Fig. 16(b). In this manner, every trace used to build the histograms in Fig. 15 was analyzed, and the resulting histograms for true one-atom contacts are shown in Fig. 17(a-c) at three different amplitudes (0.1, 0.2, 0.3 nm peak-to-peak, respectively). In Fig. 17(a-c), the red trace represents counts in the vicinity of $1G_o$ arising from two-atom contacts, and the dotted line is the cumulative of all counts akin to a traditional histogram. Notice the abrupt drop in counts in histograms for true one-atom contacts (blue trace) at conductance values exceeding $1G_o$; the slight overshoot in histograms for one-atom contacts is due to the allowance for noise in the conductance signal. Figure 17(c) shows that with increasing amplitude, the count contribution to the first peak from a single-atom contact becomes negligible (in other words, a true single-atom contact rarely forms at high amplitudes). This observed trend also explains the observed shift of peaks in Fig. 15(b) to higher conductance values because at higher amplitudes, most of the contributions to the peak at $1G_o$ come from two-atom contacts. Figure 17(d) shows the profile of the entire histogram after size-sorting to higher $G_o$ values.

Next, consider the mechanical susceptibility of the contacts (defined as $\Delta G/\Delta d$; see Fig. 18). For these measurements, the profile of the tip oscillation is shown in Fig. 18(a). The plateaus of zero amplitude between tip oscillations help determine $\Delta G$ and also ensure in subsequent analysis that the contact had reversibly returned back to its initial state after each excitation cycle, as shown in Fig. 18(b); the subscripts 'C' and 'T' represent contact in compression (piezo elongation) and tensions (piezo retraction), respectively. Figure 19(a) shows a prototypical example of a single-atom contact with near perfect transmission, highlighting saturation at $1G_o$. For such contacts, Fig. 19(b) shows that $\Delta G$ is virtually unchanged under compression, and represents the limiting case of $\Delta G/\Delta d$ approaching zero. In contrast, Figure 19(c) shows a prototypical example of a two-atom contact whose conductance changes continuously across $1G_o$ with piezo elongation and retraction, giving a finite value of $\Delta G/\Delta d$, as shown in Fig. 19(d). The resulting plot of $\Delta G/\Delta d$ in the vicinity of $1G_o$ is shown in Fig. 20(a); the inset in Fig. 20(a) shows the standard deviation of the data for single-atom contacts. A salient feature of this plot is the asymptotic drop in mechanical susceptibility to zero for single-atom contacts, and may be taken as another manifestation of the observed channel saturation. Another salient feature of Fig. 20(a) is the distinct jump in mechanical susceptibility upon transition from single-atom contacts to two-atom contacts; Fig. 20(b) shows similar behavior for larger sized contacts.

## IV. CONCLUSIONS

Using mechanical perturbation, the present study shows the hallmark of conductance quantization for a gold single-atom constriction, namely, saturation at $1G_o$ due to the availability of only a single channel for conductance for monovalent gold. The asymptotic drop in mechanical susceptibility to zero is another manifestation of channel saturation. Notwithstanding the discreteness of metallic constrictions, the finite elasticity of sufficiently stable constrictions allows for small continuous change in conductance and the ability to observe channel saturation.

There exists a multitude of atomic configurations that give rise to quasi-continuous values of conductance in a histogram. A large number of atomic configurations with similar conductance values gives rise to peaks in the conductance histogram. They tell us nothing about the stability of the atomic configurations, nor constitute a proof of conductance quantization. Position of the peaks and counts at each peak can be varied by changing the experimental conditions. Conductance traces below $1G_o$ cannot necessarily be assumed to arise from single-atom constrictions from an inspection of conductance traces alone. A two-atom constriction could also give rise to net conductance close to or even below $1G_o$. The mechanical perturbation approach offers a way to identify a true single atom constriction.

The monotony of scrutinizing the characteristics of each trace cannot be substituted by a non-discriminate statistics on a dataset, however large.

**Acknowledgments**

This work was supported by the National Science Foundation, Grant No. DMR-0706074, and this support is gratefully acknowledged. *Corresponding author: H.D.C.; E-mail: hchopra@buffalo.edu.

**FIGURE CAPTIONS:**

**FIG. 1.** (a-c) Three example conductance traces at room temperature for gold, where the last few conductance plateaus mainly take integer or near-integer values of $G_o$ prior to being ruptured.

**FIG. 2.** (a-c) Three example conductance traces at room temperature for gold, where the last few conductance plateaus take various non-integer values of $G_o$ prior to being ruptured.

**FIG. 3.** A series of gold conductance histograms at room temperature, where successive histograms represent the collective total of all preceding experimental cycles (from 400 to 5000). The inset shows the zoom-in view of the histograms corresponding to the first three peaks. The retraction speed of the piezo is 1 nm/s.

**FIG. 4.** Simultaneous measurement of (a) conductance and (b) force during the rupture of a gold constriction at room temperature. The retraction speed of the piezo is 5 nm/s.

**FIG. 5.** (a) A zoom-in view of the conductance trace shown in Fig. 2(c), along with (b) the simultaneously measured force-elongation curve. The trace reveals a small stepwise change in conductance from $1.17G_o$ to $1.02G_o$, which registers a corresponding stepwise change in force. The dotted vertical lines selectively highlight correlation between various steps in conductance and force. The inset shows a zoom-in view of the conductance trace at $\sim 1.02G_o$. The retraction speed of the piezo is 5 nm/s.

**FIG. 6.** (a) A zoom-in view of the conductance trace shown in Fig. 1(c), along with (b) the simultaneously measured force-elongation curve. The trace reveals small stepwise changes in conductance from $1.05G_o$ to $0.99G_o$, to $0.84G_o$ each of which registers a corresponding stepwise change in force. The retraction speed of the piezo is 5 nm/s.

**FIG. 7.** (a) A conductance trace whose last conductance plateau is consistently below $1G_o$. (b) Its simultaneously measured force-elongation curve. Inset in (a) shows details of the conductance trace and force-elongation curve for the plateau below $1G_o$. The retraction speed of the piezo is 5 nm/s.

**FIG. 8.** (a) Conductance trace whose last conductance plateau is consistently below $1G_o$. (b) Its simultaneously measured force-elongation curve. The retraction speed of the piezo is 5 nm/s.

**FIG. 9.** (a) An example of a conductance trace that undergo random telegraphic noise in conductance, which registers a corresponding change in force, as shown in its simultaneously measured force-elongation curve in (b). The window emphasizes the correlation between RTN and corresponding change in force. The retraction speed of the piezo is 5 nm/s.

**FIG. 10.** (a) An example of a conductance trace and (b) its corresponding force-elongation curve. The window highlights high frequency RTN at the plateau just above $1G_o$. The retraction speed of the piezo is 5 nm/s.

**FIG. 11.** (a) An example of a two-level RTN and the corresponding changes in force. The dotted lines in (a) selectively highlight the close correlation between change in conductance and measured force corresponding to RTN. (b) An example of RTN that undergoes transition between more than two levels. The two (solid) windows in (b) highlight the close correlation between change in conductance and measured force corresponding to the complex RTN transitions. In the examples shown in (a) and (b), the piezo is stationary.

**FIG. 12.** (a) A conductance trace consisting of RTN between two different atomic configurations. The inset shows a zoom-in view of the two levels in RTN. The dotted lines in the inset show the bins; bins that include stepwise transition are highlighted in green. (b)

Conductance histogram of the trace shown in (a). (c) Standard deviation of all the bins from the trace in (a). The data points within the red and blue ellipses in (c) correspond to the true standard deviation of Level-1 and Level-2, respectively. All other data points in (c) are artifacts arising from bins that contain the stepwise change between Level-1 and Level-2.

**FIG. 13.** (a) A dust map showing true standard deviation versus conductance that excludes any stepwise transitions in the recorded traces. The dust map consists of over 61,000 data points. The retraction speed for individual traces is 3.5 pm/s. (b) Zoom-in view of the dust map in the vicinity of $1G_o$. The red curve in (b) is the average standard deviation of the dust map, whereas the green curve roughly outlines the lower bound of the standard deviation.

**FIG. 14.** (a) Average standard deviation of the conductance juxtaposed with (b) the histogram shown in Fig. 3 with 5000 traces.

**FIG. 15.** (a) Histograms built by superimposing various peak-to-peak amplitudes of mechanical oscillation on the retracting gold tip. Each histogram is built using 1000 traces. The inset shows the profile of the oscillations superimposed on the retracting tip. Retraction rate is 3.5 pm/s. (b) Zoom-in view of the first peak as a function of oscillation amplitude. In (a-b) count percentage (ordinate) is obtained by normalizing the counts between zero and $10G_o$ to 100%.

**Fig. 16.** (a) Continuous and reversible change in conductance below and above $1G_o$ as the piezo is elongated and retracted for a two-atom contact whose net conductance is close to $1G_o$. (b) Saturation at $1G_o$ due to the availability of only a single channel for conductance across a single gold atom as the hallmark of conductance quantization. (c-d) Change in conductance versus elongation/retraction corresponding to (a-b), respectively. For (c) six contiguous cycles are plotted whereas for (d) three contiguous cycles are plotted.

**FIG. 17.** Conductance histograms for true one-atom contacts (blue trace) at three different amplitudes (a) 0.1 nm, (b) 0.2 nm, (c) 0.3 nm peak-to-peak. The red trace corresponds to two-atom histogram in the vicinity of $1G_o$. The dotted trace is the cumulative total of all counts in the histogram. (d) Split conductance histograms to higher $G_o$ values. For atomic configurations greater than a single-atom, the total counts may not be definitive owning to the possibility of, for example, a three-atom configuration having a conductance much less than $2G_o$.

**FIG. 18.** (a) A profile of the tip oscillation for mechanical susceptibility ($\Delta G/\Delta d$) measurements of the contacts. (b) Example of a corresponding conductance response to the oscillation profile in (a).

**FIG. 19.** (a) A prototypical example of a single-atom contact with near perfect transmission, highlighting saturation at $1G_o$. (b) Corresponding conductance versus tip oscillation trace. (c) A prototypical example of a two-atom contact with net conductance close to $1G_o$. (d) Corresponding conductance versus tip oscillation trace.

**FIG. 20.** (a) A plot of mechanical susceptibility ($\Delta G/\Delta d$) versus conductance in the vicinity of $1G_o$ (blue solid dots for true one-atom contacts and red open dots for two-atom contacts). Inset shows standard deviation of the data for single-atom contacts. Note the asymptotic drop in $\Delta G/\Delta d$ to zero for single-atom contacts, and a large jump in mechanical susceptibility for two-atom contacts. (b) Similar behavior for larger sized contacts. Total number of measurements is ~2000.

**Reference:**

[1] N. Oncel, Journal of Physics: Condensed Matter **20**, 393001 (2008).

[2] Y. Selzer and D. L. Allara, Annual Review of Physical Chemistry **57**, 593 (2006).

[3] F. Grill and F. Moresco, Journal of Physics: Condensed Matter **2006**, S1887 (2006).

[4] F. Moresco, Physics Reports **399**, 175 (2004).

[5] J. Kröger, N. Néel, and L. Limot, Journal of Physics: Condensed Matter **20**, 222301 (2008).

[6] A. Sperl, J. Kröger, and R. Berndt, physica status solidi (b) **247**, 1077 (2010).

[7] A. Calzolari, C. Cavazzoni, and M. Buongiorno Nardelli, Physical Review Letters **93**, 096404 (2004).

[8] S. N. Behera, S. Gayen, G. V. Ravi Prasad, and S. M. Bose, Physica B: Condensed Matter **390**, 124 (2007).

[9] S. Wippermann, N. Koch, and W. G. Schmidt, Physical Review Letters **100**, 106802 (2008).

[10] S. Díaz-Tendero, S. Fölsch, F. E. Olsson, A. G. Borisov, and J.-P. Gauyacq, Nano Letters **8**, 2712 (2008).

[11] A. Sperl, J. Kröger, N. Néel, H. Jensen, R. Berndt, A. Franke, and E. Pehlke, Physical Review B **77**, 085422 (2008).

[12] N. Nilius, T. M. Wallis, and W. Ho, Science **297**, 1853 (2002).

[13] J. Lagoute, X. Liu, and S. Fölsch, Physical Review B **74**, 125410 (2006).

[14] K. Schouteden, E. Lijnen, D. A. Muzychenko, A. Ceulemans, L. F. Chibotaru, P. Lievens, and C. Van Haesendonck, Nanotechnology **20**, 395401 (2009).

[15] S. L. Elizondo and J. W. Mintmire, International Journal of Quantum Chemistry **105**, 772 (2005).

[16] Y. Kihira, T. Shimada, Y. Matsuo, E. Nakamura, and T. Hasegawa, Nano Letters **9**, 1442 (2009).

[17] J. Park, et al., Nature **417**, 722 (2002).

[18] B. Xu and N. J. Tao, Science **301**, 1221 (2003).

[19] X. L. Li, S. Z. Hua, H. D. Chopra, and N. J. Tao, Micro & Nano Letters, IET **1**, 83 (2006).

[20] D. G. Cahill, W. K. Ford, K. E. Goodson, G. D. Mahan, A. Majumdar, H. J. Maris, R. Merlin, and S. R. Phillpot, Journal of Applied Physics **93**, 793 (2003).

[21] M. Tsutsui, S. Kurokawa, and A. Sakai, Applied Physics Letters **90**, 133121 (2007).

[22] A. Sperl, J. Kröger, and R. Berndt, Physical Review B **81**, 035406 (2010).

[23] Z. Huang, F. Chen, R. D'Agosta, P. A. Bennett, M. Di Ventra, and N. Tao, Nat Nano **2**, 698 (2007).

[24] D. M. Eigler and E. K. Schweizer, Nature **344**, 524 (1990).

[25] L. Bartels, G. Meyer, and K. H. Rieder, Physical Review Letters **79**, 697 (1997).

[26] A. Deshpande, H. Yildirim, A. Kara, D. P. Acharya, J. Vaughn, T. S. Rahman, and S. W. Hla, Physical Review Letters **98**, 028304 (2007).

[27] S.-W. Hla, K.-F. Braun, and K.-H. Rieder, Physical Review B **67**, 201402 (2003).

[28] L. Gross, F. Mohn, N. Moll, P. Liljeroth, and G. Meyer, Science **325**, 1110 (2009).

[29] O. Custance, R. Perez, and S. Morita, Nat Nano **4**, 803 (2009).

[30] M. Ternes, C. P. Lutz, C. F. Hirjibehedin, F. J. Giessibl, and A. J. Heinrich, Science **319**, 1066 (2008).

[31] Y. Sugimoto, P. Pou, O. Custance, P. Jelinek, M. Abe, R. Perez, and S. Morita, Science **322**, 413 (2008).

[32] U. Landman, W. D. Luedtke, N. A. Burnham, and R. J. Colton, Science **248**, 454 (1990).

[33] A. P. Sutton and J. B. Pethica, Journal of Physics: Condensed Matter **2**, 5317 (1990).

[34] S. Ciraci, E. Tekman, A. Baratoff, and I. P. Batra, Physical Review B **46**, 10411 (1992).

[35] P. E. Marszalek, W. J. Greenleaf, H. Li, A. F. Oberhauser, and J. M. Fernandez, Proceedings of the National Academy of Sciences of the United States of America **97**, 6282 (2000).

[36] G. Rubio-Bollinger, S. R. Bahn, N. Agraït, K. W. Jacobsen, and S. Vieira, Physical Review Letters **87**, 026101 (2001).

[37] G. Rubio, N. Agraït, and S. Vieira, Physical Review Letters **76**, 2302 (1996).

[38] B. J. van Wees, H. van Houten, C. W. J. Beenakker, J. G. Williamson, L. P. Kouwenhoven, D. van der Marel, and C. T. Foxon, Physical Review Letters **60**, 848 (1988).

[39] D. A. Wharam, et al., Journal of Physics C: Solid State Physics **21**, L209 (1988).

[40] L. Olesen, E. Laegsgaard, I. Stensgaard, F. Besenbacher, J. Schiøtz, P. Stoltze, K. W. Jacobsen, and J. K. Nørskov, Physical Review Letters **72**, 2251 (1994).

[41] L. Olesen, E. Laegsgaard, I. Stensgaard, F. Besenbacher, J. Schiøtz, P. Stoltze, K. W. Jacobsen, and J. K. Nørskov, Physical Review Letters **74**, 2147 (1995).

[42] Z. Gai, Y. He, H. Yu, and W. S. Yang, Physical Review B **53**, 1042 (1996).

[43] J. M. Krans, C. J. Muller, I. K. Yanson, T. C. M. Govaert, R. Hesper, and J. M. van Ruitenbeek, Physical Review B **48**, 14721 (1993).

[44] J. M. Krans, J. M. van Ruitenbeek, V. V. Fisun, I. K. Yanson, and L. J. de Jongh, Nature **375**, 767 (1995).

[45] A. I. Yanson, Ph.D. Thesis, Universiteit Leiden, The Netherlands, 2001, p. 81.

[46] A. I. Yanson and J. M. van Ruitenbeek, Physical Review Letters **79**, 2157 (1997).

[47] E. Scheer, N. Agrait, J. C. Cuevas, A. L. Yeyati, B. Ludoph, A. Martin-Rodero, G. R. Bollinger, J. M. van Ruitenbeek, and C. Urbina, Nature **394**, 154 (1998).

[48] Y. Kurui, Y. Oshima, M. Okamoto, and K. Takayanagi, Physical Review B **77**, 161403 (2008).

[49] Y. Kurui, Y. Oshima, M. Okamoto, and K. Takayanagi, Physical Review B **79**, 165414 (2009).

[50] M. D. Huntington, J. N. Armstrong, M. R. Sullivan, S. Z. Hua, and H. D. Chopra, Physical Review B **78**, 035442 (2008).

[51] A. I. Yanson, I. K. Yanson, and J. M. van Ruitenbeek, Nature **400**, 144 (1999).

[52] A. I. Yanson, I. K. Yanson, and J. M. van Ruitenbeek, Physical Review Letters **84**, 5832 (2000).

[53] A. I. Yanson, I. K. Yanson, and J. M. van Ruitenbeek, Physical Review Letters **87**, 216805 (2001).

[54] T. N. Todorov and A. P. Sutton, Physical Review Letters **70**, 2138 (1993).

[55] T. N. Todorov and A. P. Sutton, Physical Review B **54**, R14234 (1996).

[56] M. Dreher, F. Pauly, J. Heurich, J. C. Cuevas, E. Scheer, and P. Nielaba, Physical Review B **72**, 075435 (2005).

[57] J. E. Sader, J. W. M. Chon, and P. Mulvaney, Review of Scientific Instruments **70**, 3967 (1999).

[58] The total elongation of 0.62 nm results from 0.41 nm from piezo elongation (1.45-1.04 nm) plus 0.21 nm from cantilever motion (corresponding to the stepwise changes in force). The total cantilever motion is obtained by dividing the magnitude of two force jumps in the last plateau (not including the last jump to rupture) by the spring constant of the cantilever (11 N/m or 11 nN/nm). The total of these force jumps is 2.31 nN, corresponding to cantilever motion of 0.21 nm.

[59] The total elongation of 0.425 nm results from 0.355 nm from piezo elongation (1.515-1.16 nm) plus 0.07 nm from cantilever motion (during the stepwise changes in force). The total force jump is 0.75 nN, corresponding to cantilever motion of 0.07 nm.

[60] H. Ohnishi, Y. Kondo, and K. Takayanagi, Nature **395**, 780 (1998).

[61] G. Timp, R. E. Behringer, and J. E. Cunningham, Physical Review B **42**, 9259 (1990).

[62] D. H. Cobden, N. K. Patel, M. Pepper, D. A. Ritchie, J. E. F. Frost, and G. A. C. Jones, Physical Review B **44**, 1938 (1991).

[63] D. H. Cobden, A. Savchenko, M. Pepper, N. K. Patel, D. A. Ritchie, J. E. F. Frost, and G. A. C. Jones, Physical Review Letters **69**, 502 (1992).

[64] C. Dekker, A. J. Scholten, F. Liefrink, R. Eppenga, H. van Houten, and C. T. Foxon, Physical Review Letters **66**, 2148 (1991).

[65] R. T. Wakai and D. J. V. Harlingen, Applied Physics Letters **49**, 593 (1986).

[66] R. T. Wakai and D. J. Van Harlingen, Physical Review Letters **58**, 1687 (1987).

[67] F. Liu, M. Bao, K. L. Wang, D. Zhang, and C. Zhou, Physical Review B **74**, 035438 (2006).

[68] K. S. Ralls and R. A. Buhrman, Physical Review Letters **60**, 2434 (1988).

[69] K. S. Ralls and R. A. Buhrman, Physical Review B **44**, 5800 (1991).

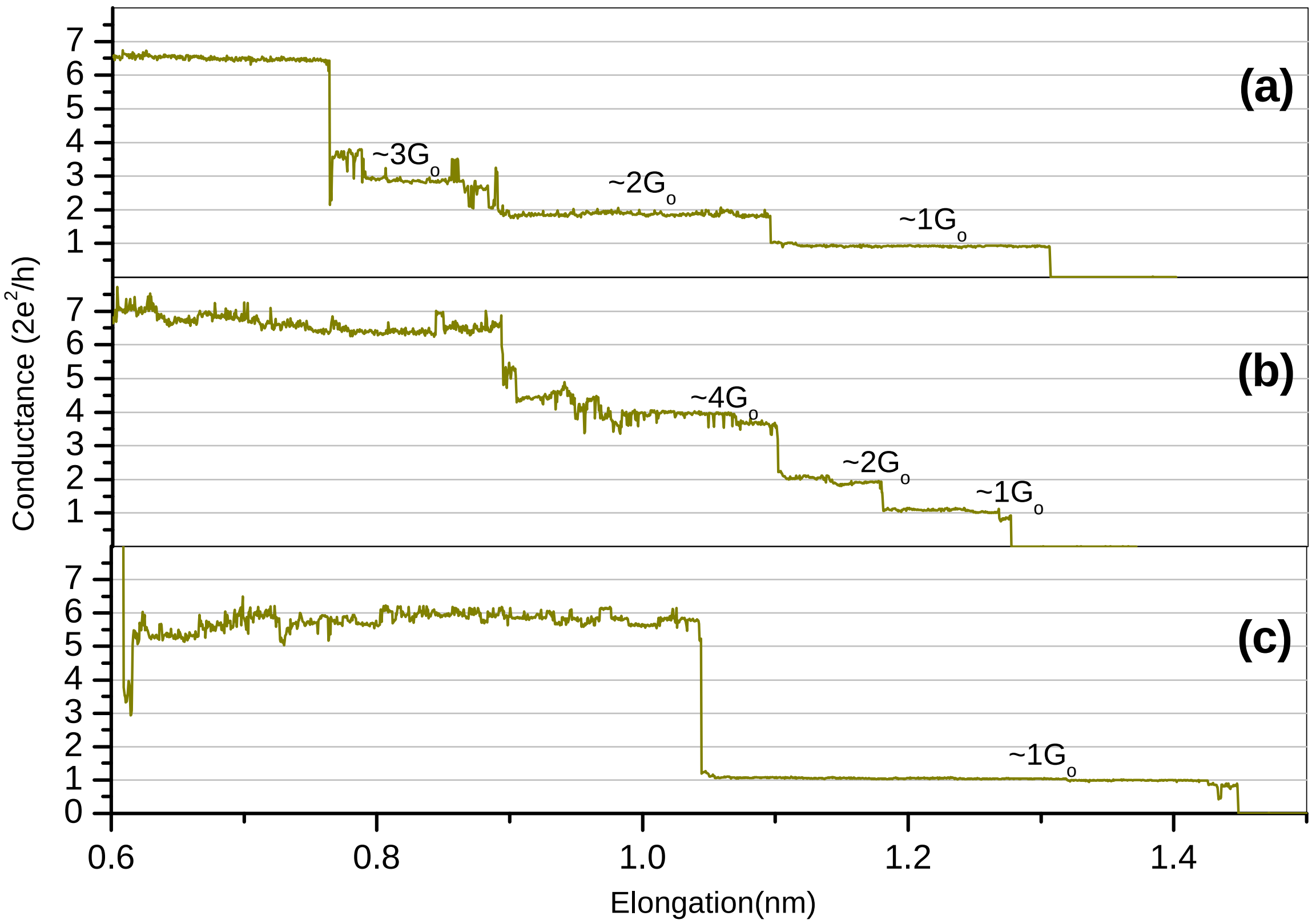

**Figure 1**

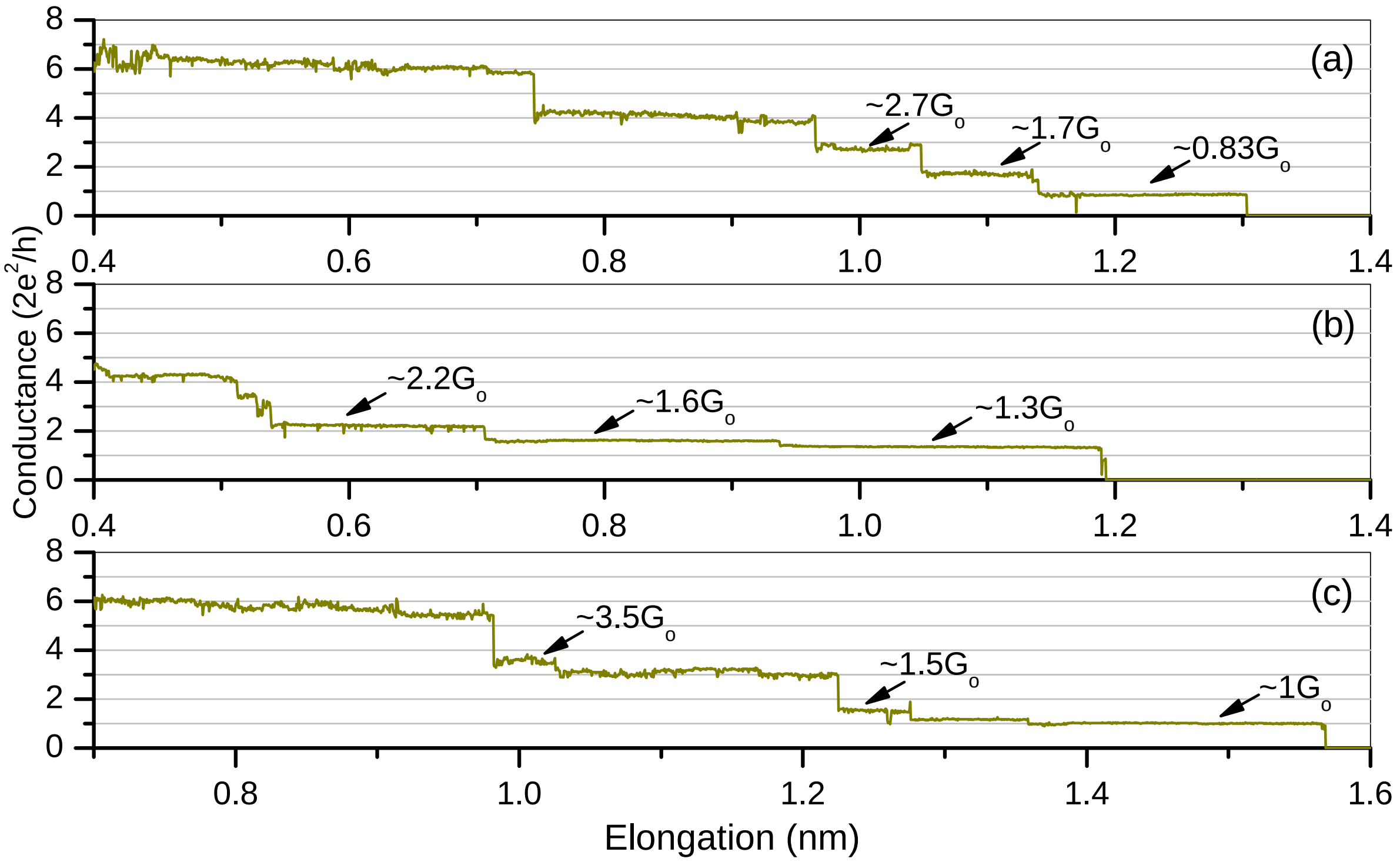

**Figure 2**

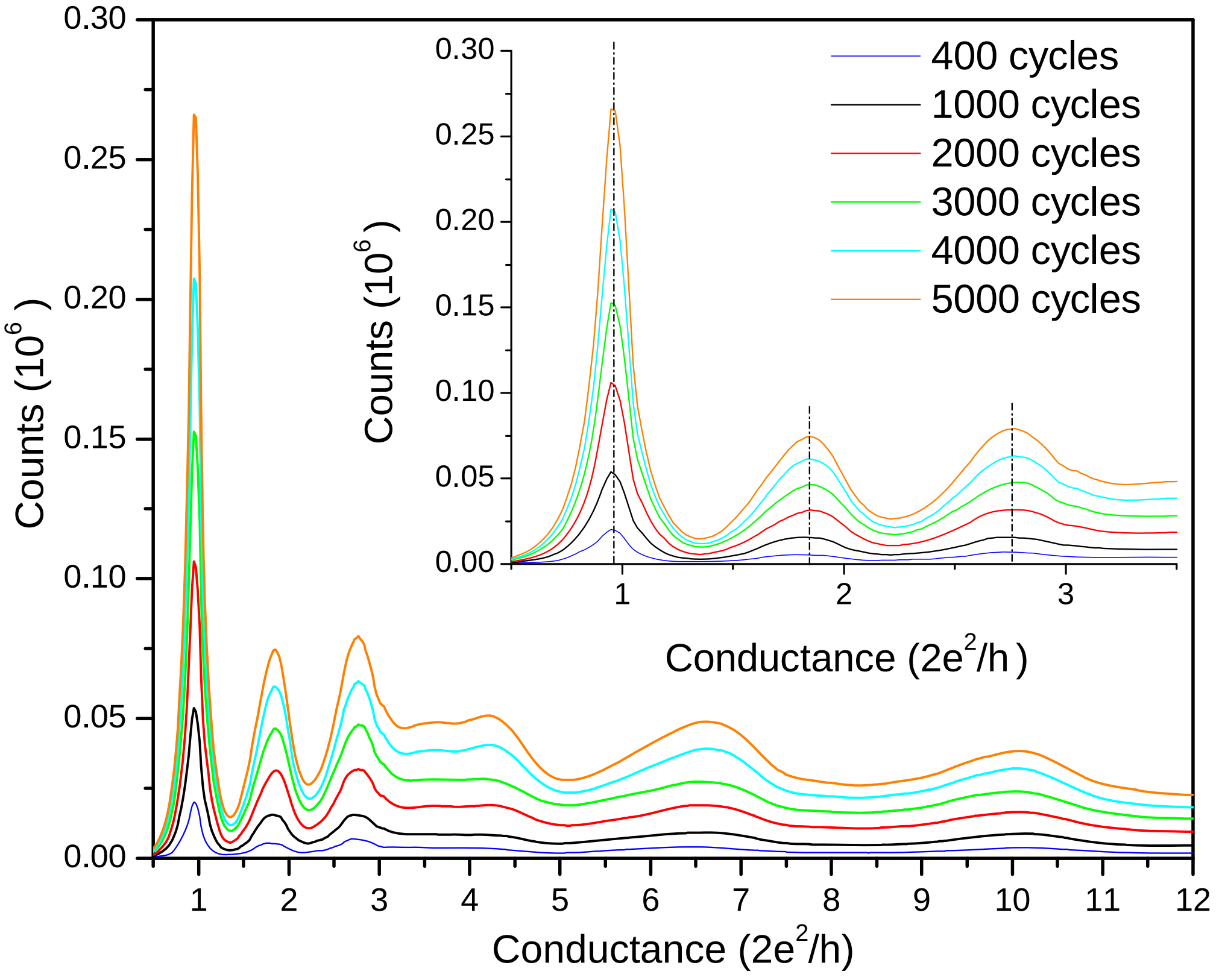

**Figure 3**

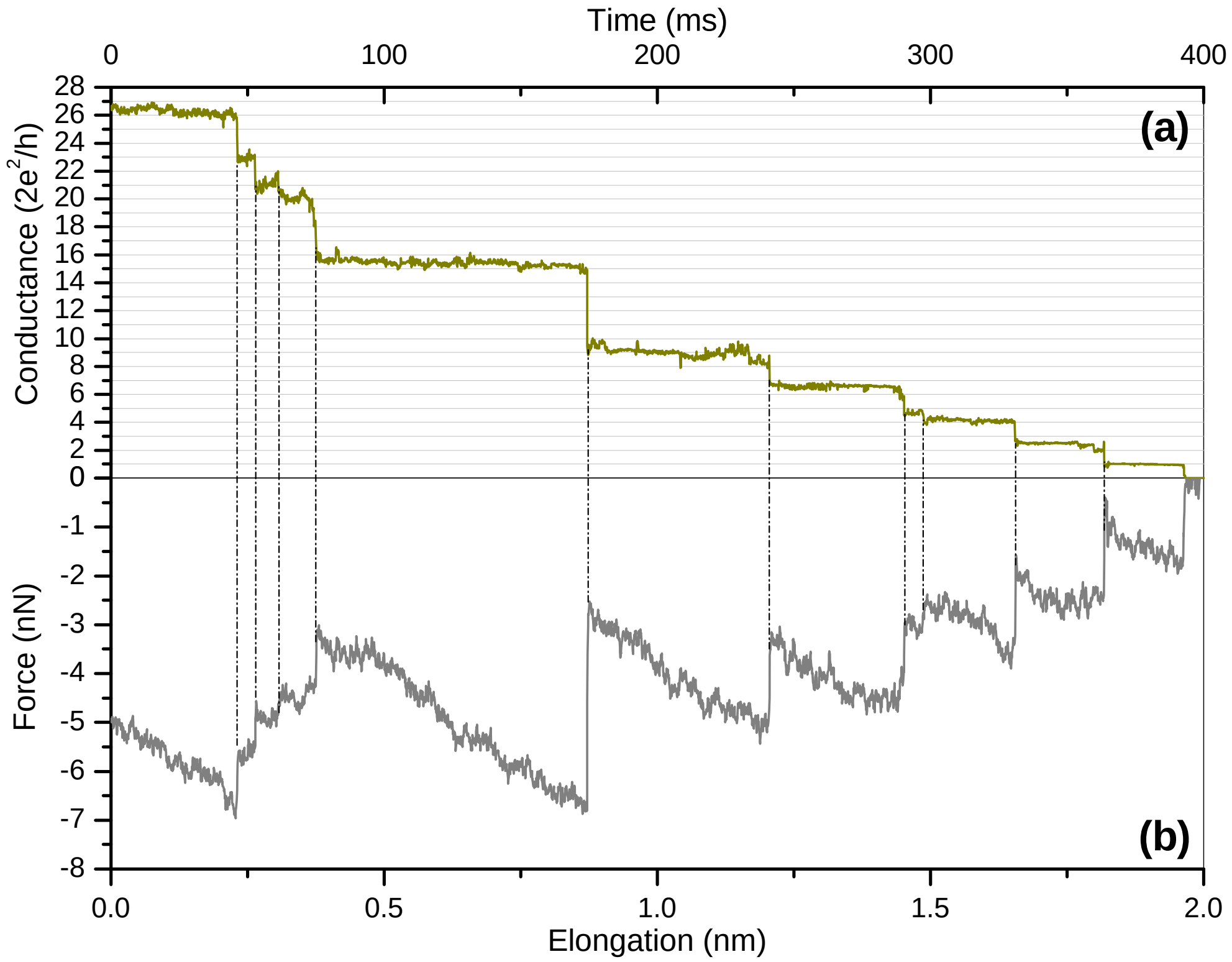

**Figure 4**

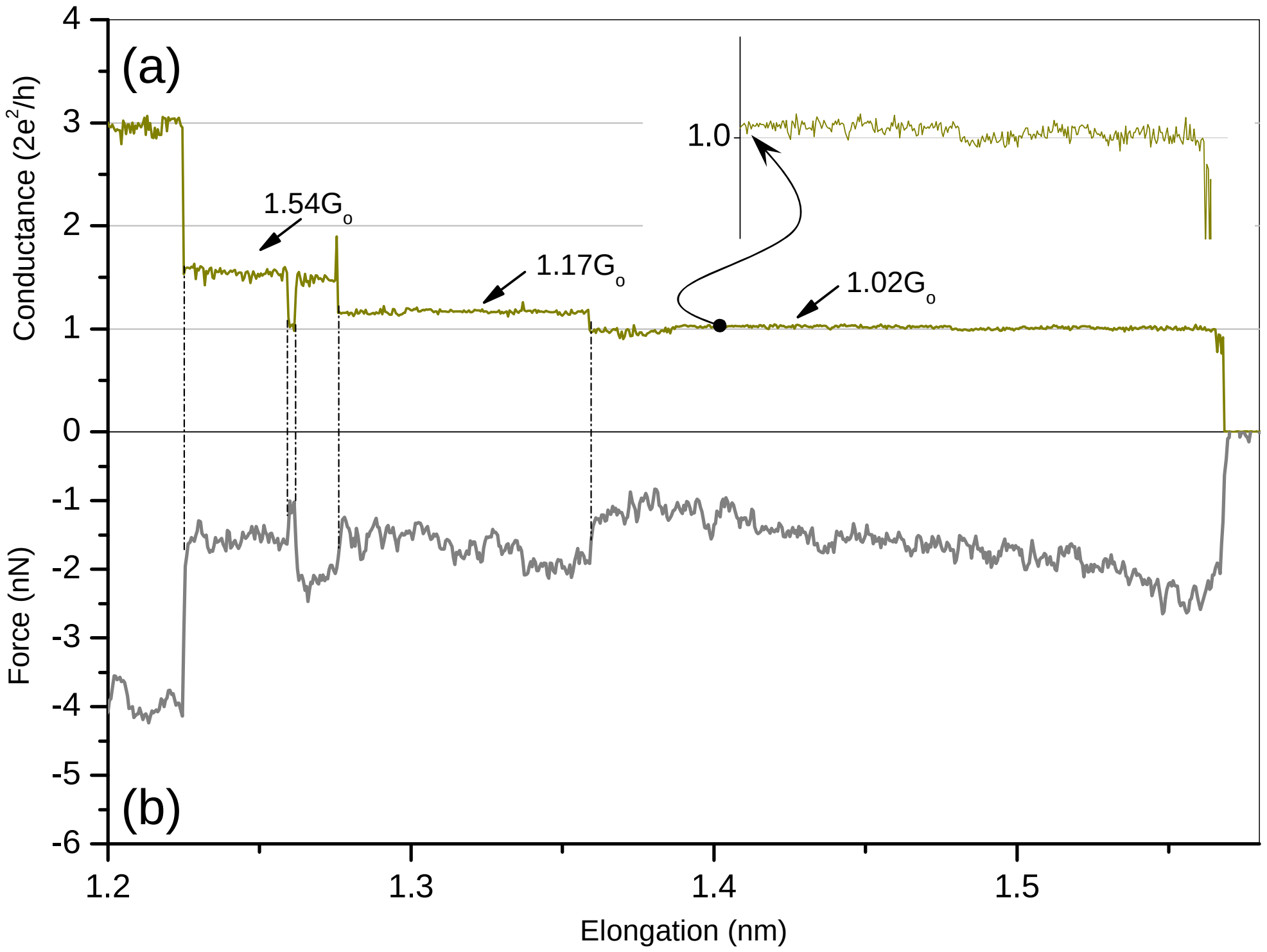

**Figure 5**

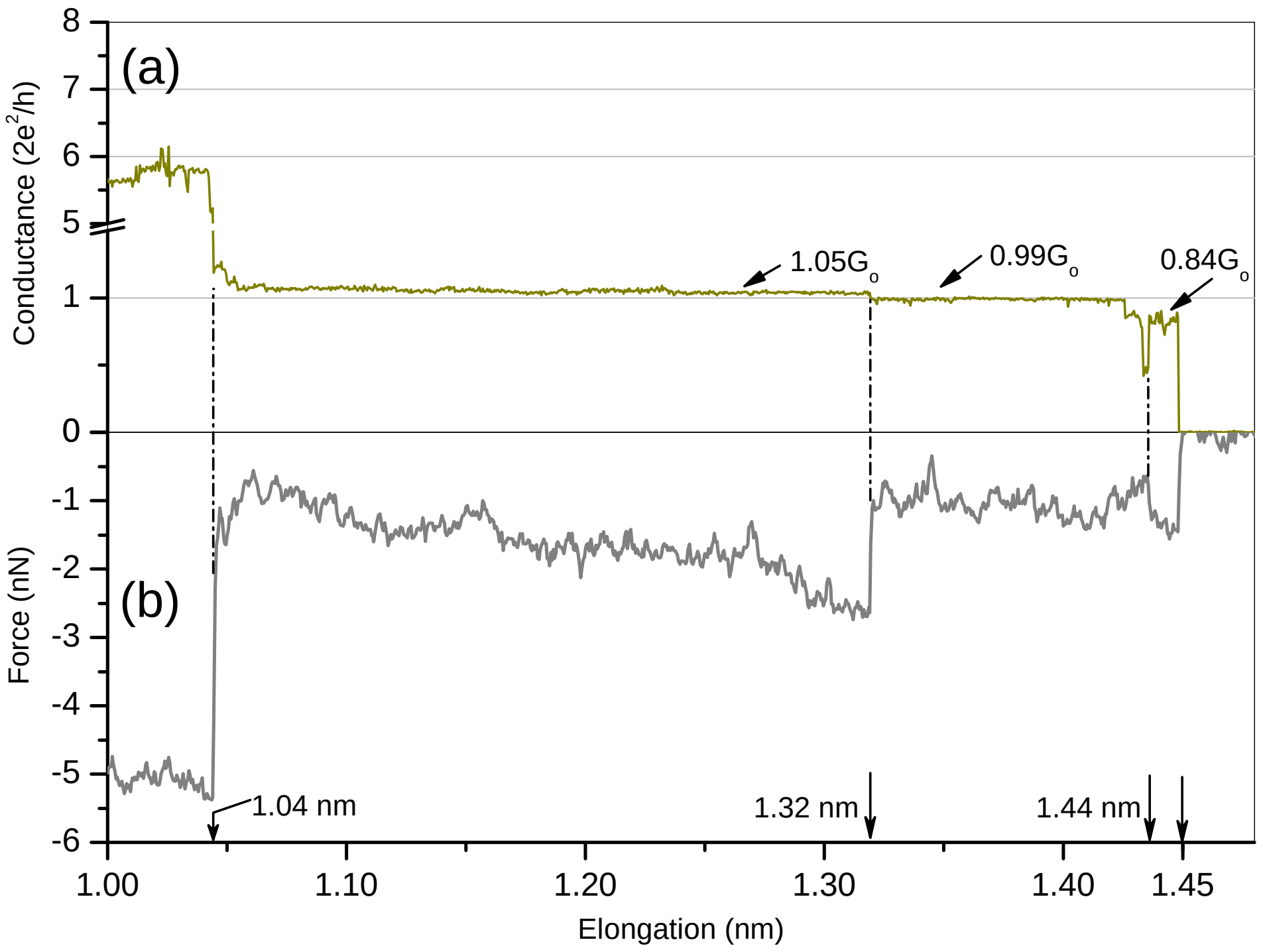

**Figure 6**

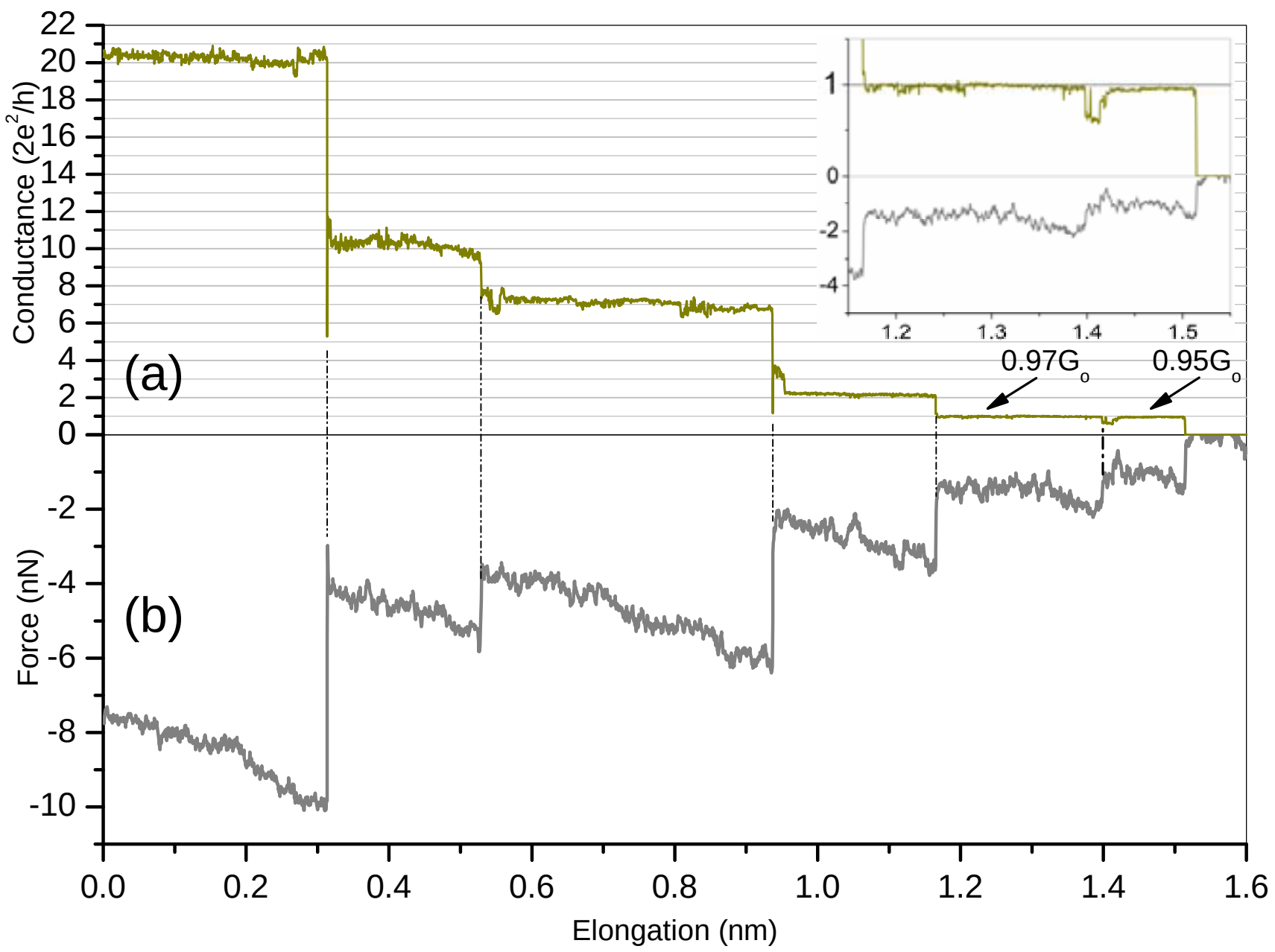

**Figure 7**

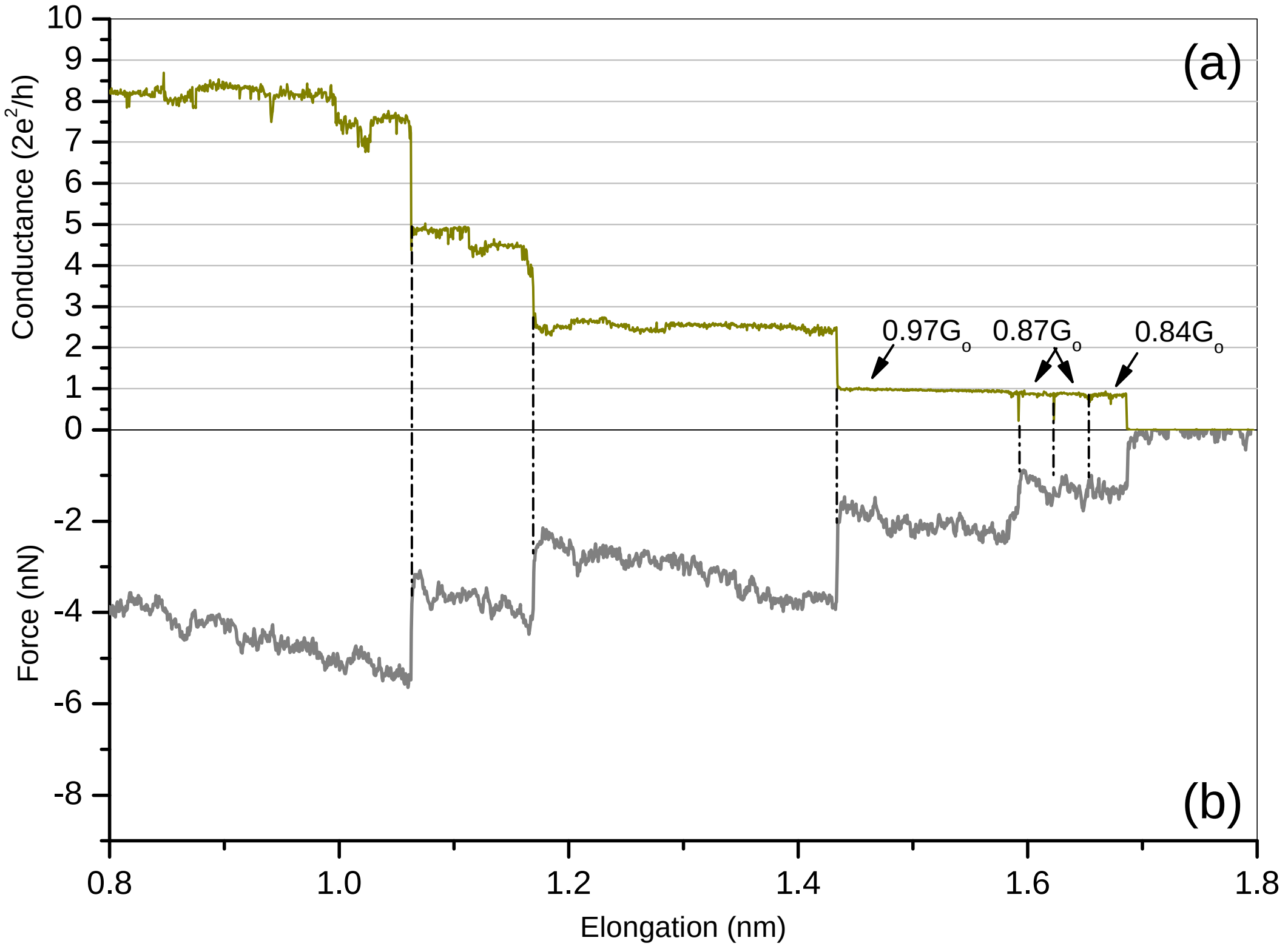

**Figure 8**

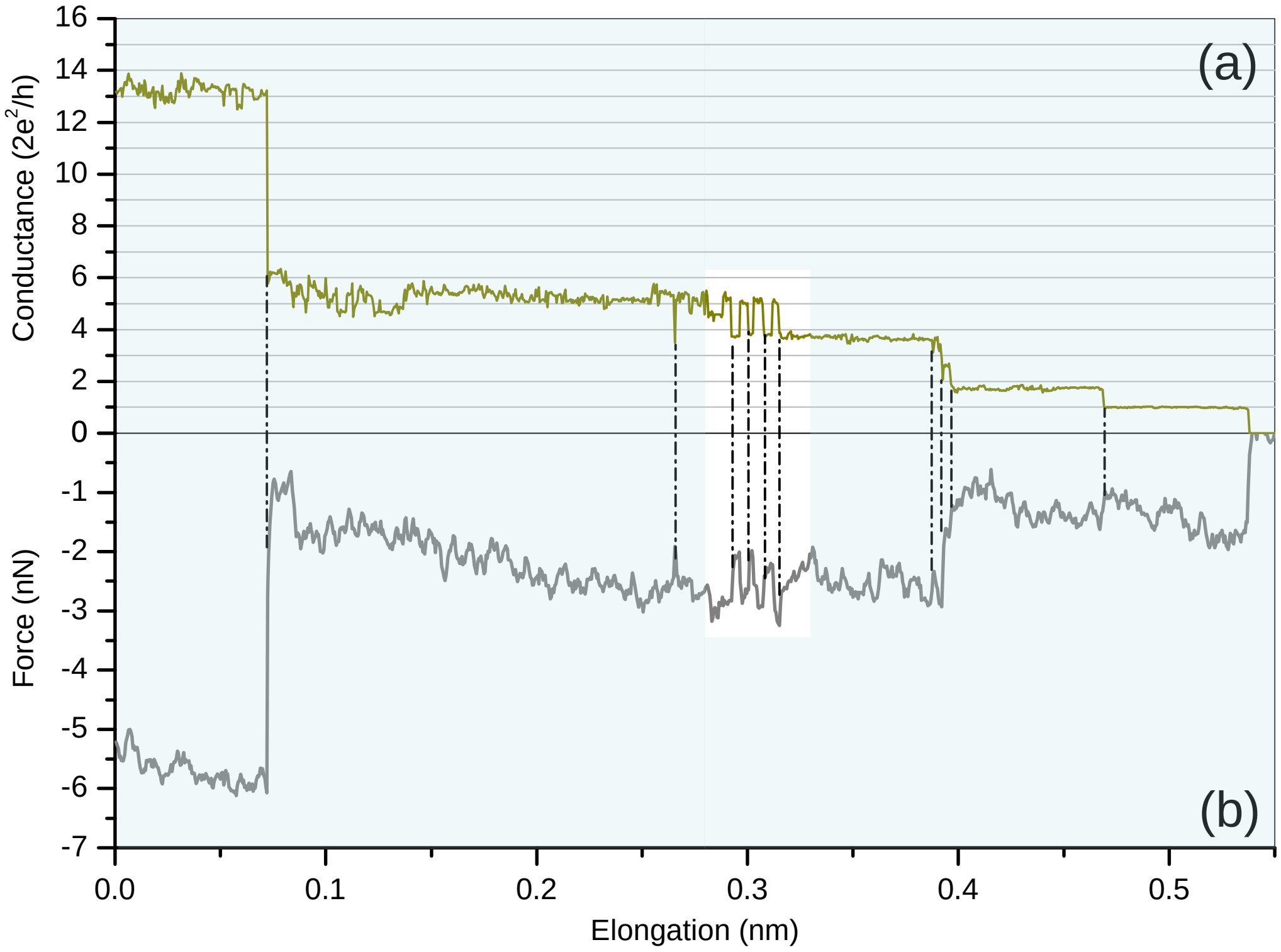

**Figure 9**

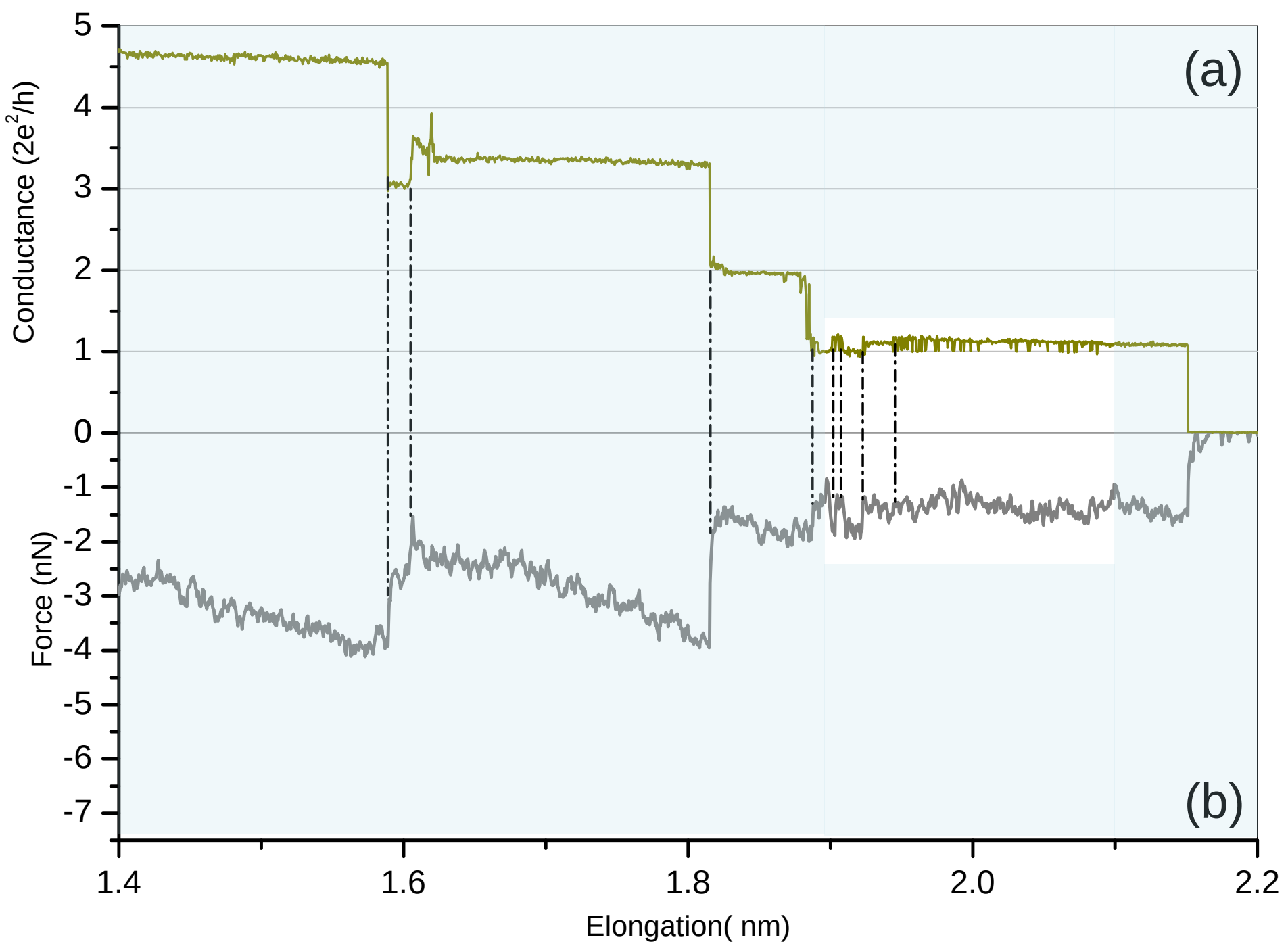

**Figure 10**

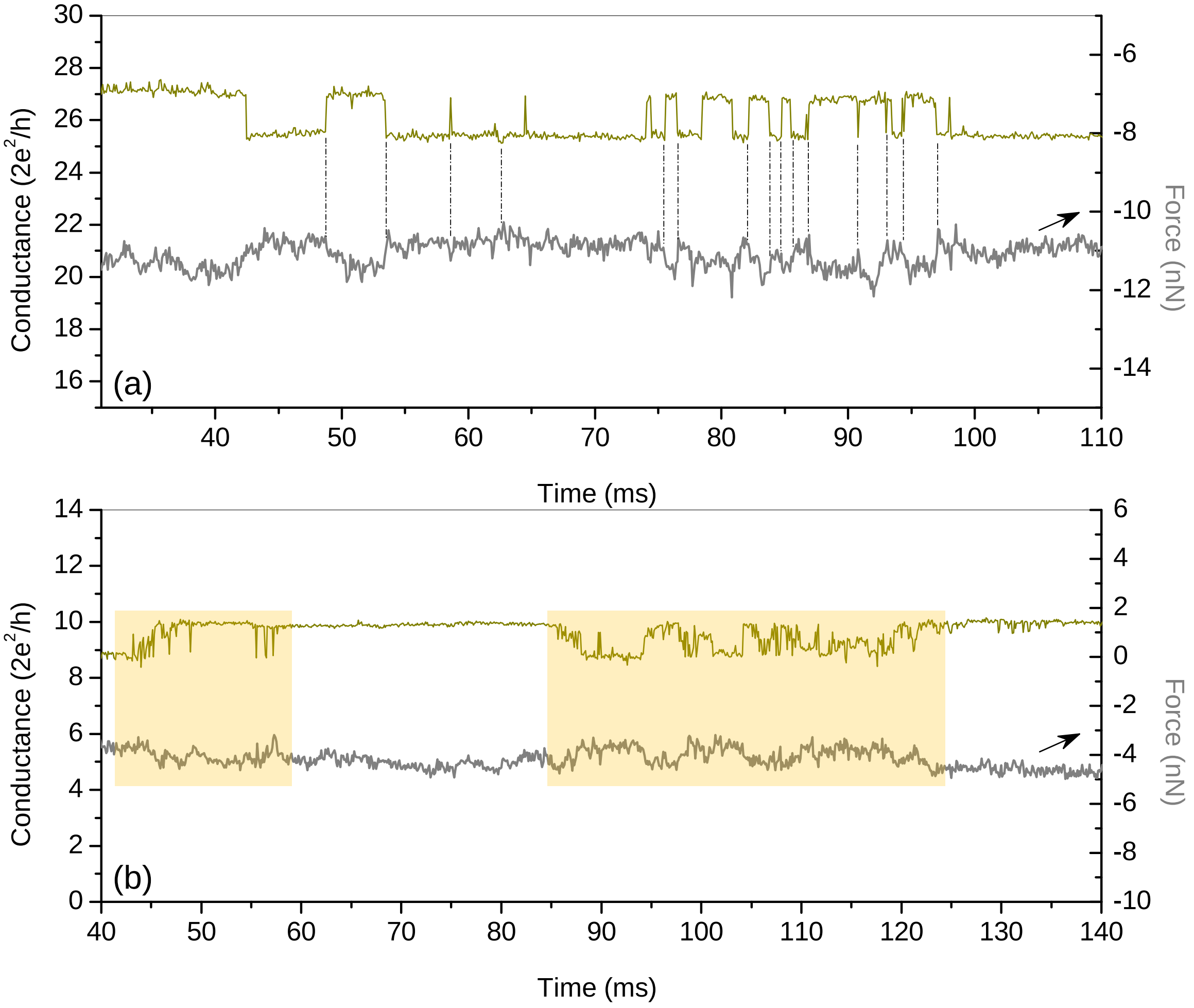

**Figure 11**

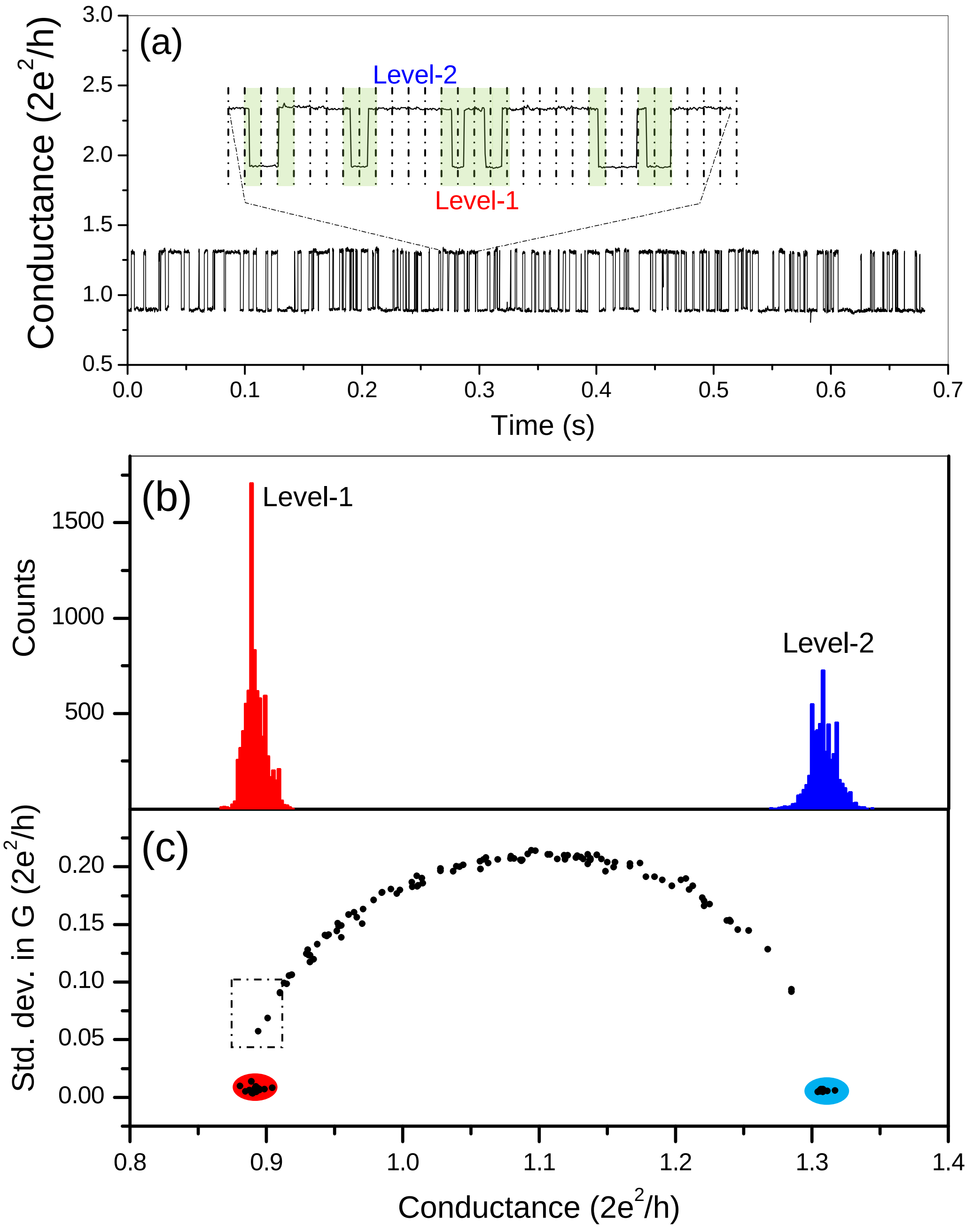

**Figure 12**

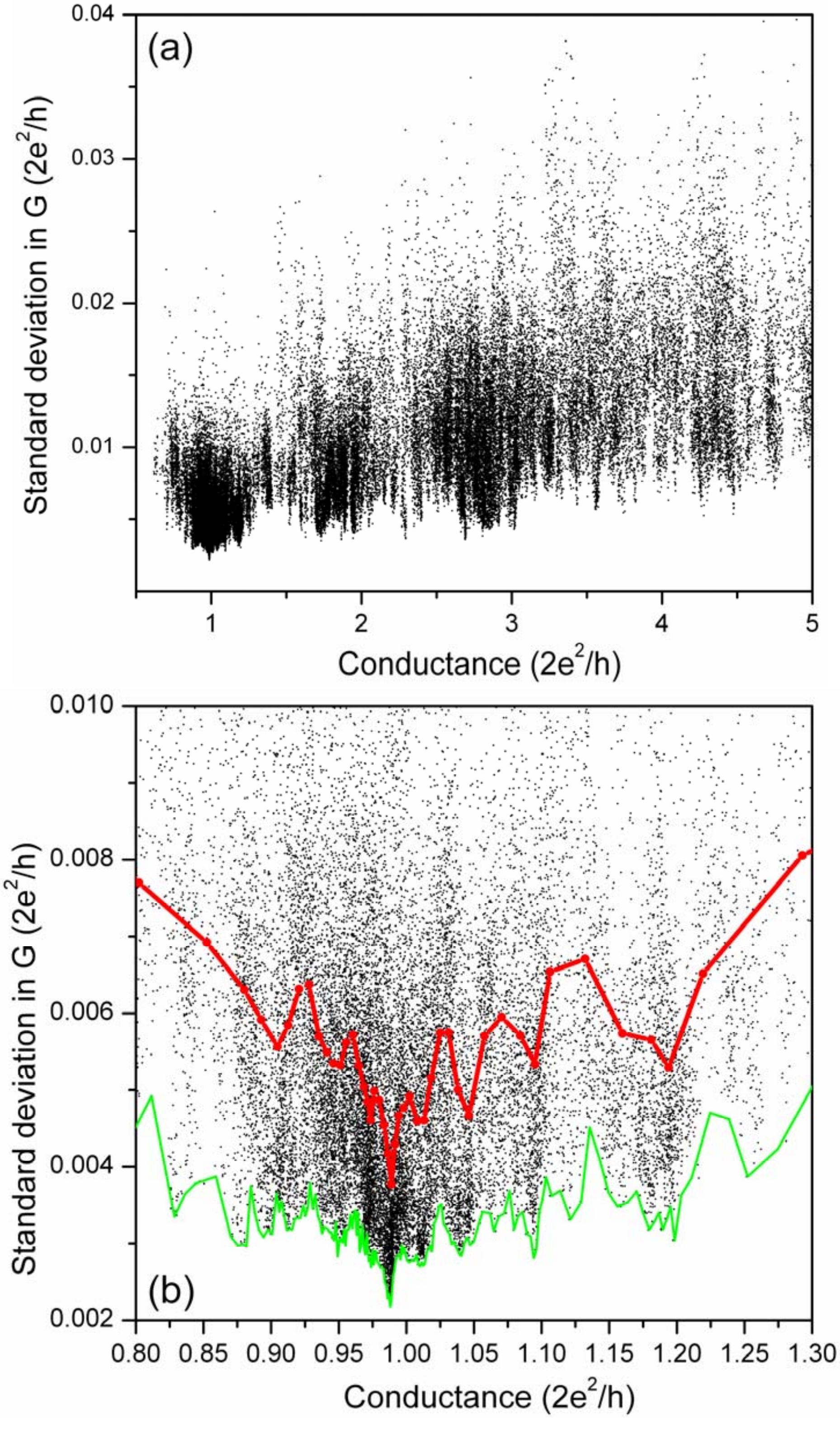

**Figure 13**

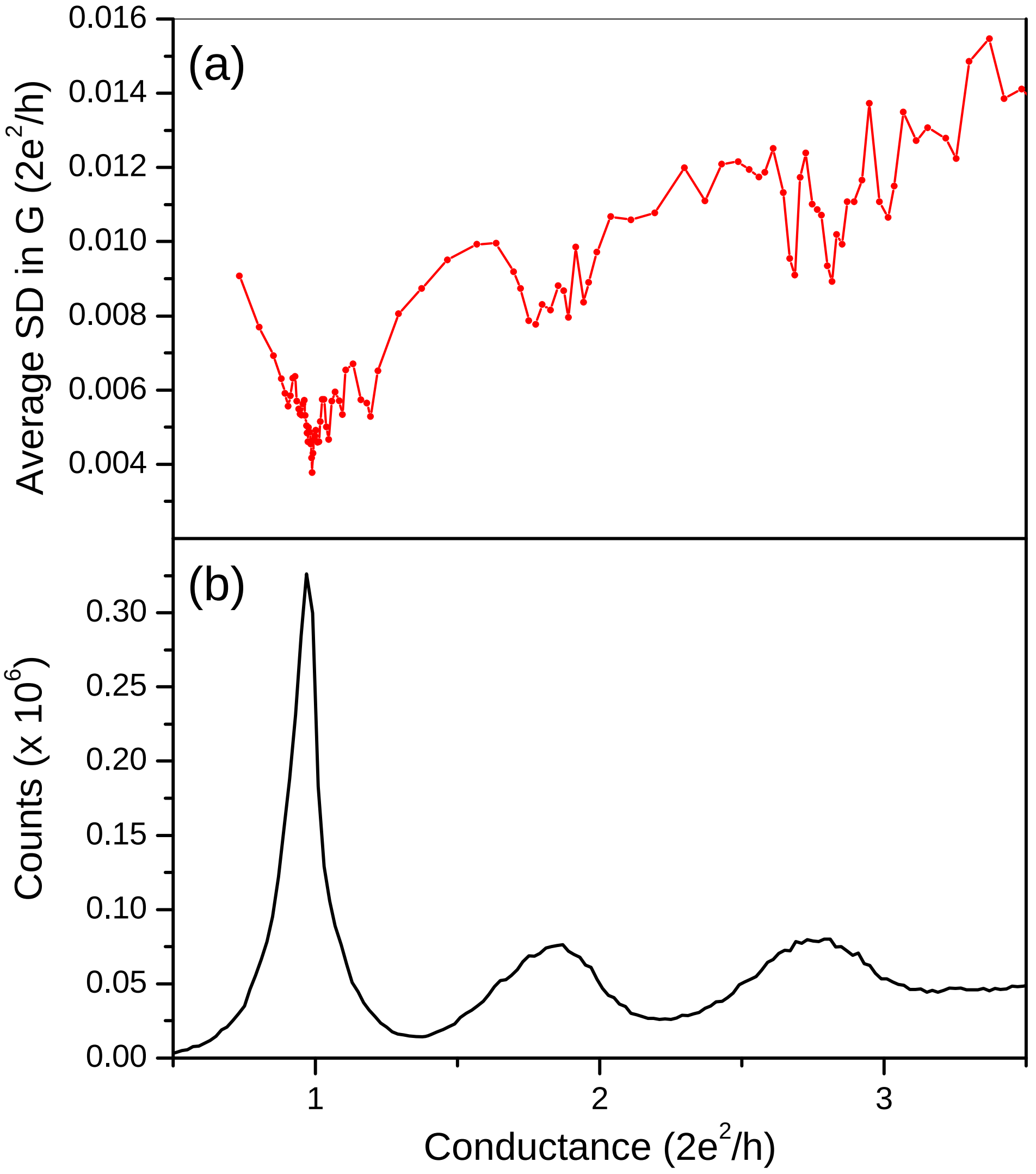

**Figure 14**

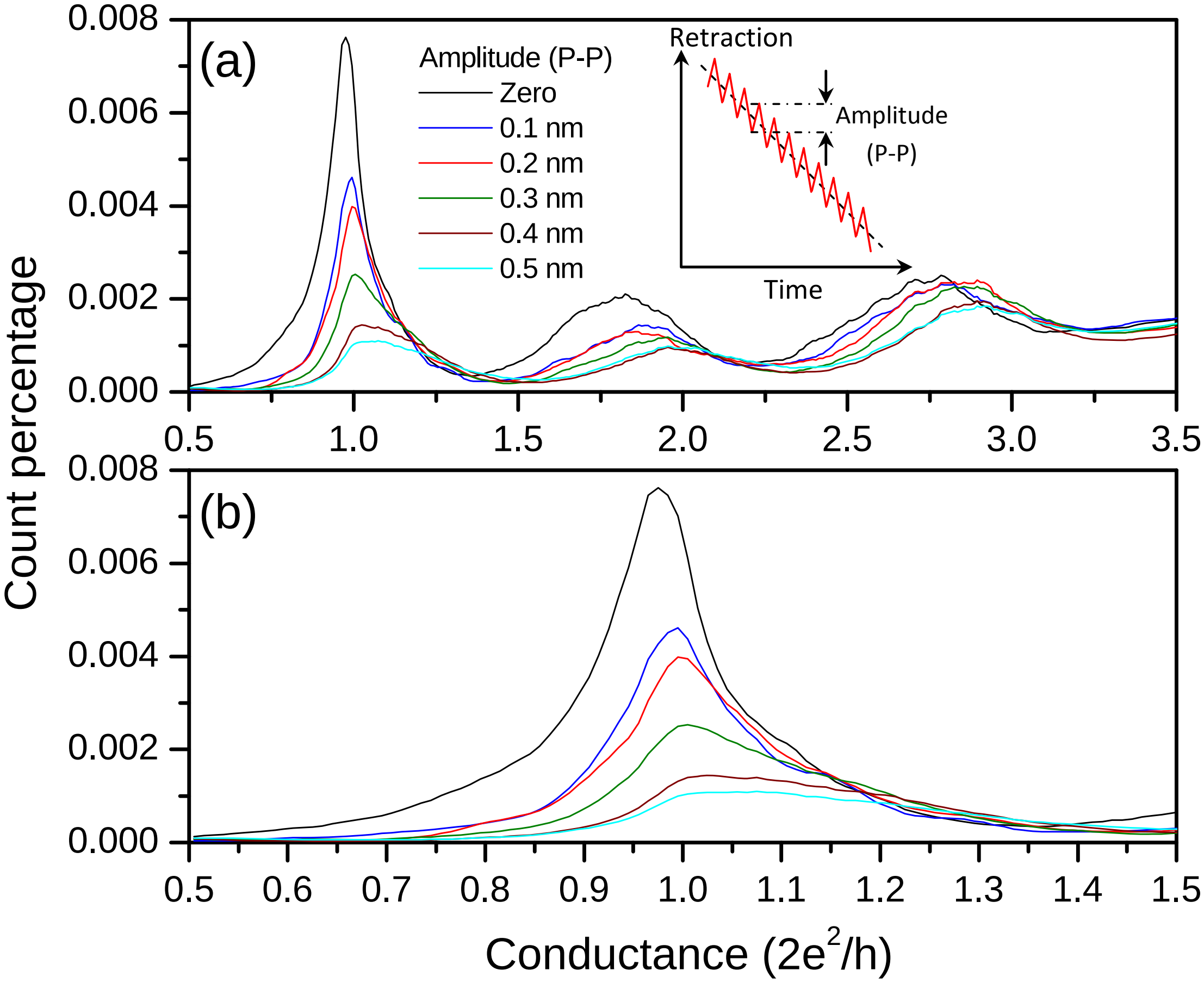

**Figure 15**

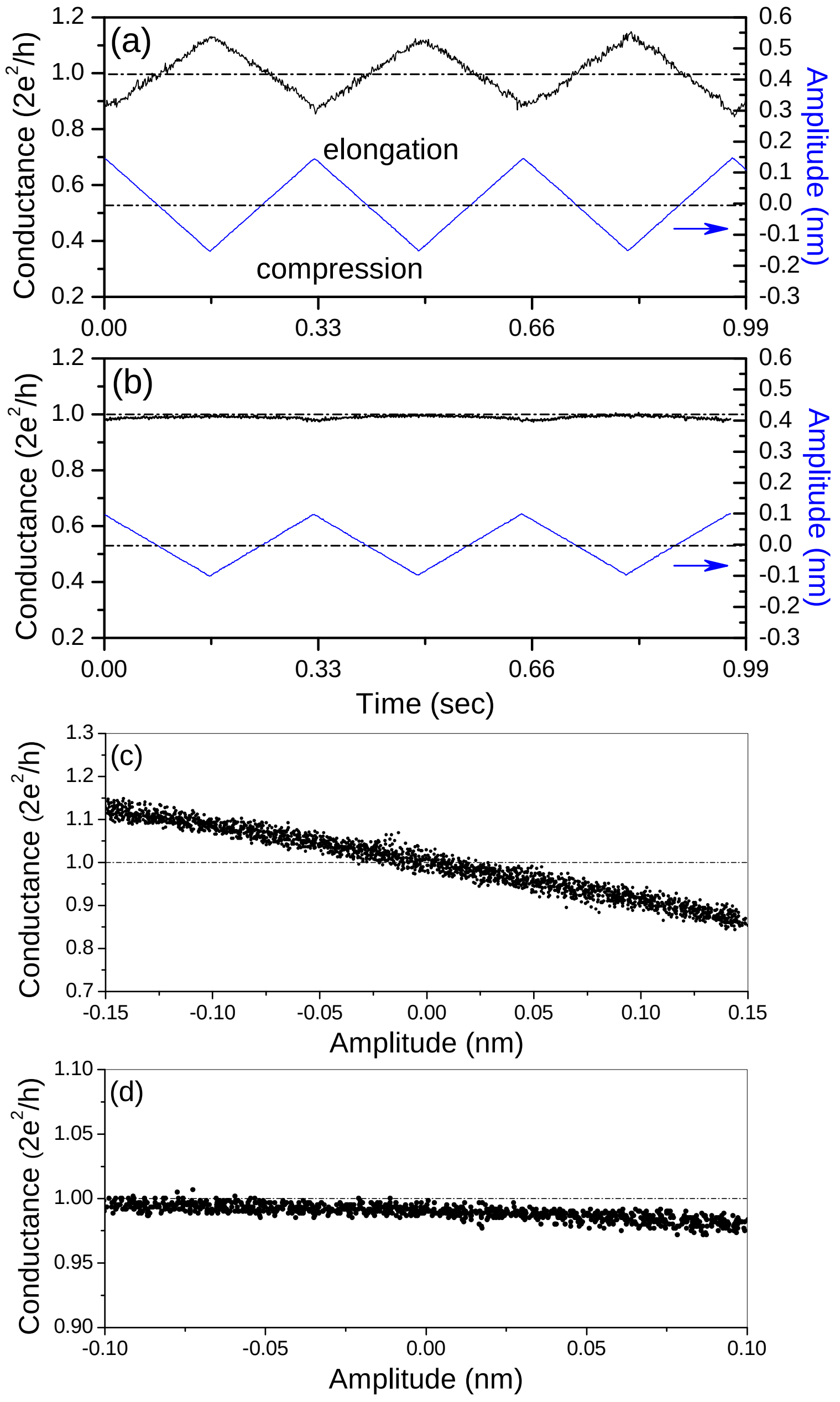

**Figure 16**

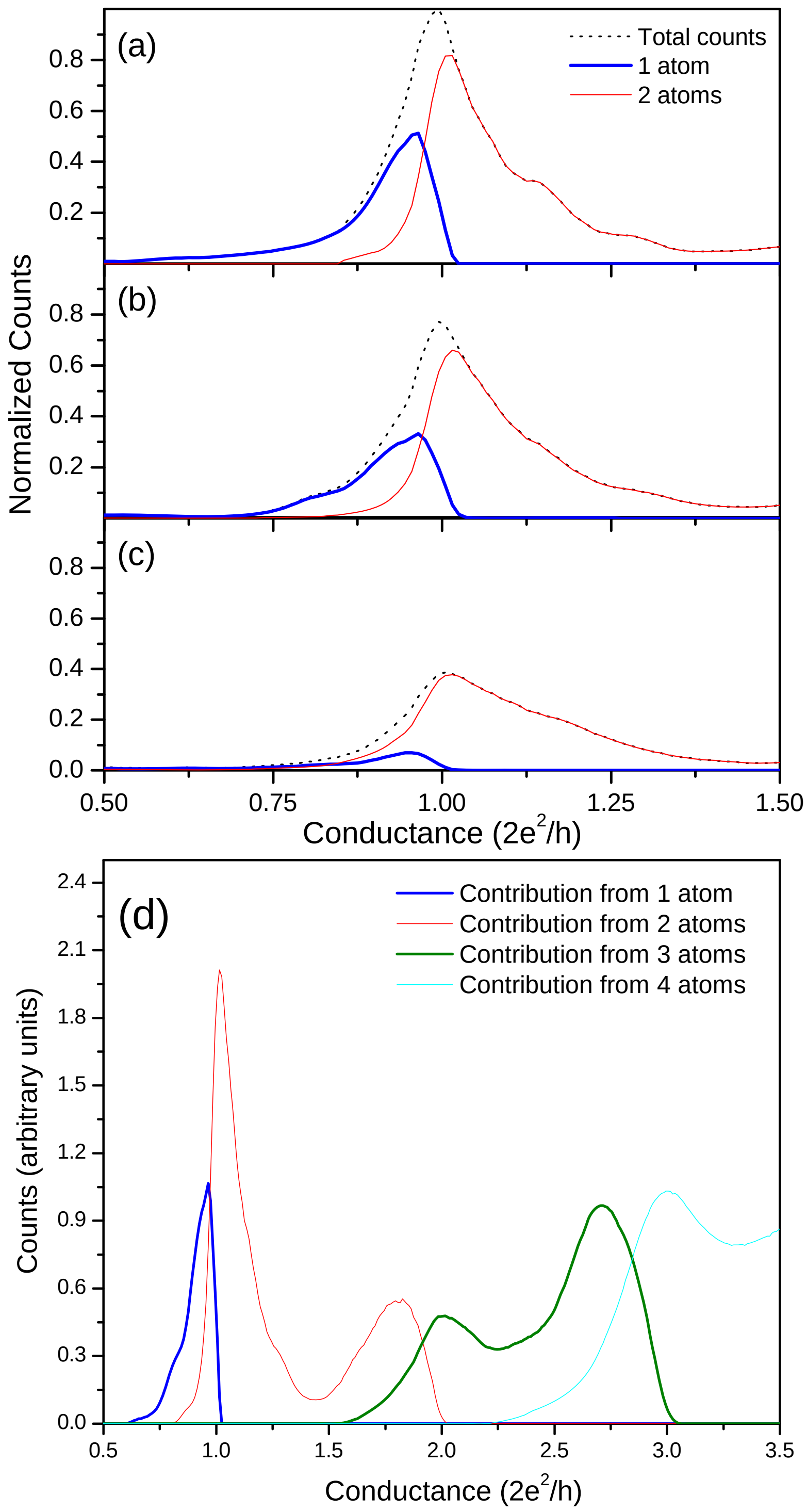

**Figure 17**

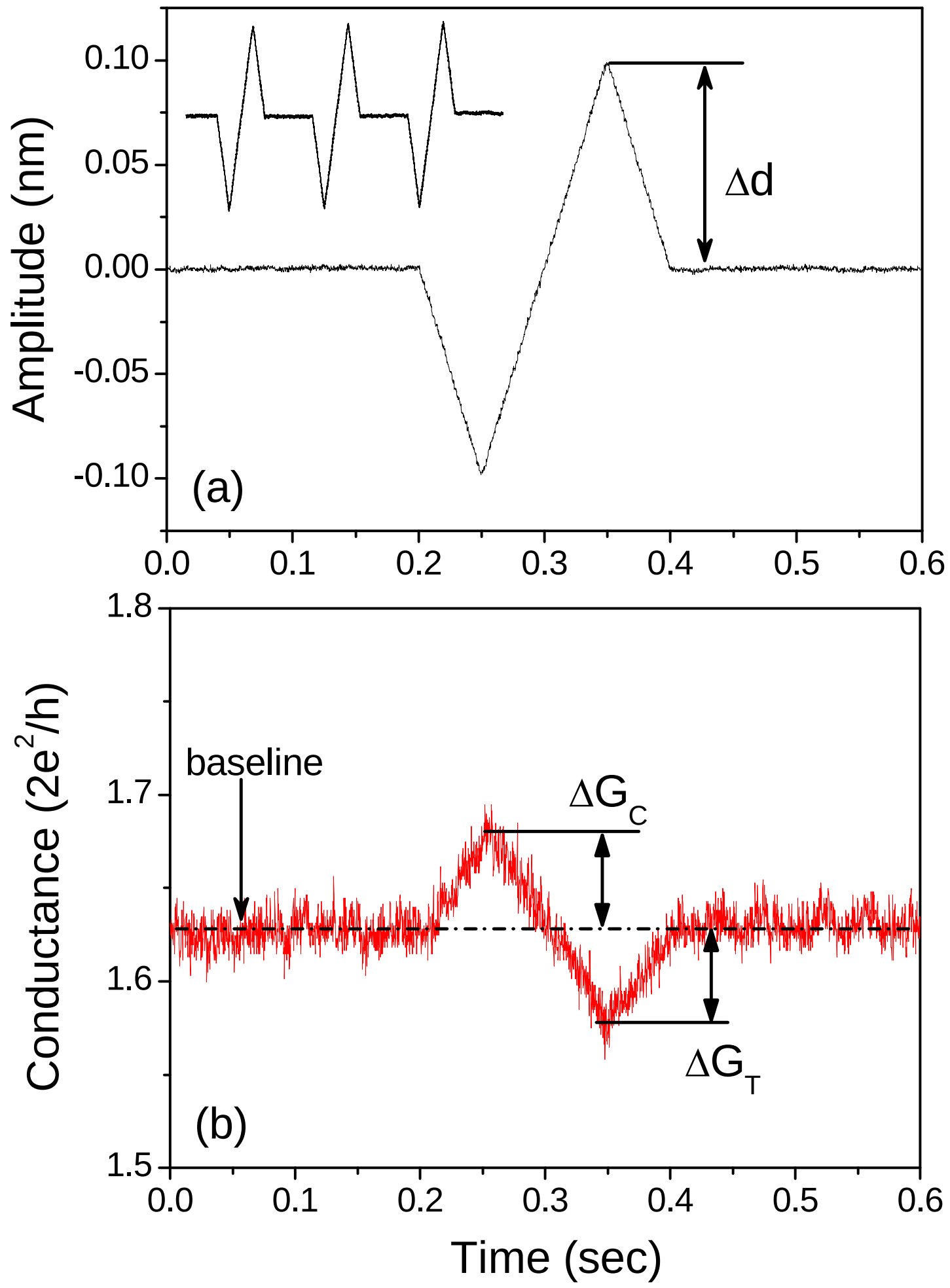

**Figure 18**

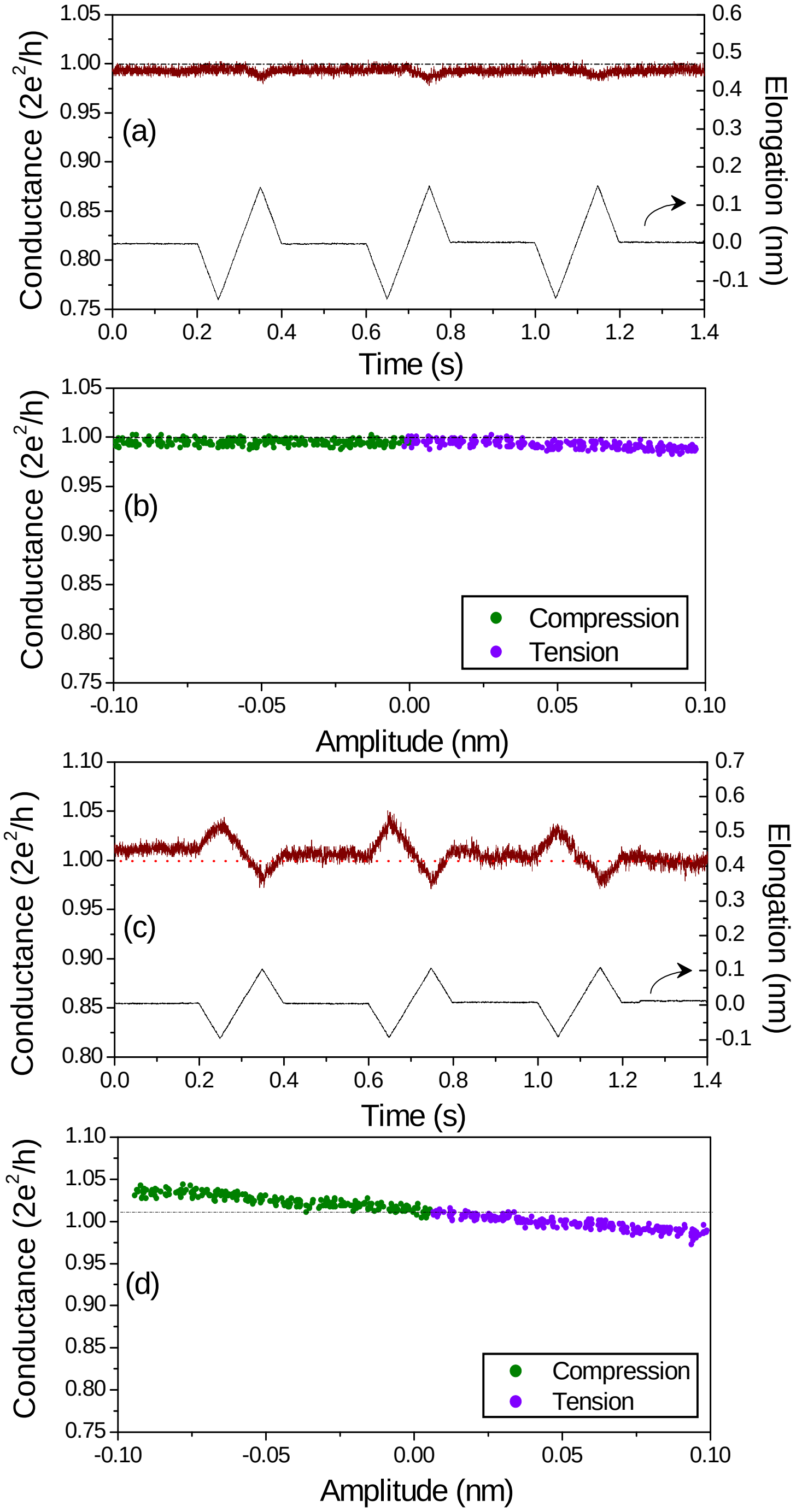

**Figure 19**

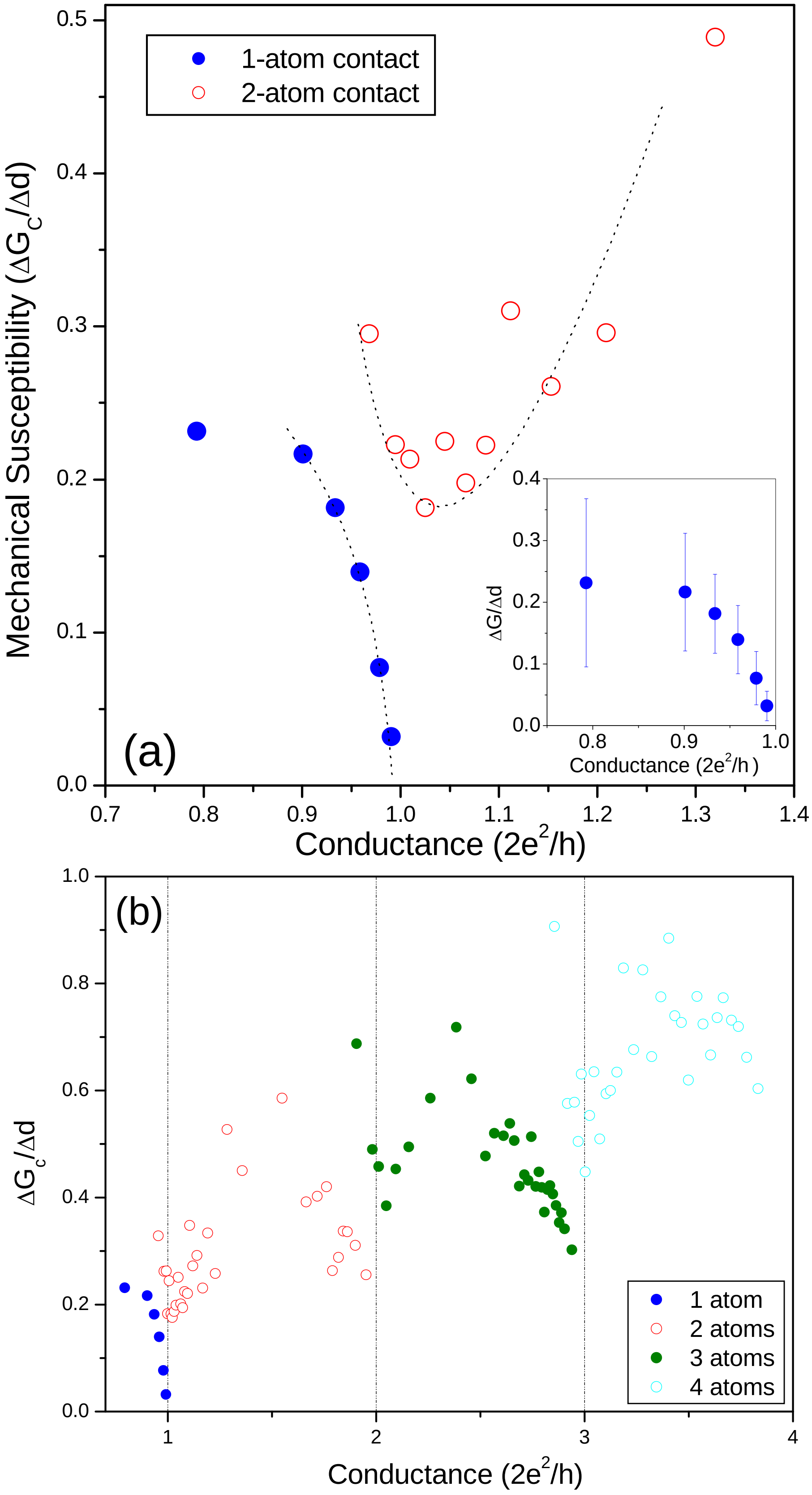

**Figure 20**